\documentclass[preprint]{aastex}
\usepackage{lscape}
\begin{document}

\title{The Evolution of Dusty Debris Disks Around Solar Type Stars}

\author{Laura Vican\altaffilmark{1} \& Adam Schneider\altaffilmark{2}}

\altaffiltext{1}{Department of Physics and Astronomy, University of California, Los Angeles, CA 90095, USA (lvican@astro.ucla.edu)}
\altaffiltext{2}{Department of Physics and Astronomy, University of Georgia, Athens, GA, 30602, USA. Current Address: Department of Physics and Astronomy, The University of Toledo, Toledo, OH, 43606, USA (aschneid@physast.uga.edu).} 

\begin{abstract}
We used chromospheric activity to determine the ages of 2,820 field stars.. We searched these stars for excess emission at 22 $\mu$m with the Wide-Field Infrared Survey Explorer. Such excess emission is indicative of a dusty debris disk around a star. We investigated how disk incidence trends with various stellar parameters, and how these parameters evolve with time. We found 22 $\mu$m excesses around 98 stars (a detection rate of 3.5$\%$). Seventy-four of these 98 excess sources are presented here for the first time. We also measured the abundance of lithium in 8 dusty stars in order to test our stellar age estimates.
\end{abstract} 

\section{INTRODUCTION}
While most of the geological evidence about the evolution of our solar system has been erased by cataclysmic events, we can study the evolution of stellar systems like our own by observing circumstellar debris disks around solar-type (F, G, and K type) stars. Much work has been done to observe and characterize debris disks (e.g. Bryden et al. 2006), but it has been notoriously complicated to track the evolution of these disks. The root of the problem is the difficulty in determining stellar age. While the ages of clusters and associations can be determined by their bulk properties (i.e. HR diagrams), such techniques are not useful for isolated field stars. 

Because of the difficulty of stellar age-dating, the study of debris disk evolution has been largely constrained to A-type stars (Su et al. 2006, Rieke et al. 2005). Since A-type stars evolve quickly on the main sequence, their ages can be estimated from stellar isochrones. Su et al. (2006) found that dust around A-type stars declines with age as t$_{0}$/t, where t$_{0}$=150 Myr. 

Isochrone dating is not adequate for solar type stars, however, since they evolve slowly on the main sequence. It is important to extend the study of debris disk evolution to solar-type stars, since they offer the best evidence about the evolution of our own solar system. Chromospheric activity dating has a well-calibrated age relation, and carries smaller errors than isochrone dating (Mamajek $\&$ Hillenbrand 2008). We constructed a sample of 2,820 main-sequence field stars for which we have calculated age based on chromospheric activity. By using field stars, we created a sample with a smooth background age distribution. Thus, any dependence of debris disk incidence on stellar age should stand out.

The Wide-Field Infrared Survey Explorer (WISE; Wright et al. 2010) offers a unique opportunity to discover new circumstellar disks. WISE Band 4 (22 $\mu$m, hereafter W4) can trace infrared emission from the small (micron-sized) dust grains which dominate the emission from debris disks. Not only is WISE sensitive to 6 mJy (5$\sigma$) at 22 $\mu$m (Wright et al. 2010), but it also has the potential to catch debris disks around stars toward which other infrared observatories such as the Spitzer Space Telescope and Herschel Space Observatory may not have pointed. By pairing new age determination techniques with the all-sky coverage of WISE, we are able to provide new insight into the evolution of debris disks around solar-type stars.

Bryden et al. (2006) used Spitzer to search for IR excess emission around 127 F, G, and K type stars. They found seven stars with excess at 70 $\mu$m and only one star with excess at 24 $\mu$m. Trilling et al. (2008) followed by observing 184 F, G, and K type stars with Spitzer, finding seven with 24 $\mu$m excesses (an excess detection rate of 3.8$\%$). Spangler et al. (2001) observed $\sim$150 pre-main-sequence and main-sequence stars (mostly in clusters) with the Infrared Space Observatory (ISO). These were mostly young ($<$1 Gyr) F and G type stars. Thus, their detection rate will be higher than in an unbiased survey. They found 33 stars with evidence of IR excess (a detection rate of 22 $\%$). Koerner et al. 2010 used Spitzer to search for debris disks among 634 solar-type stars, finding a detection rate of 4.6$\%$ at 24 $\mu$m and 4.8$\%$ at 70 $\mu$m. In the present paper, we consider a sample of  2,820 stars with activity-determined ages which we examined with WISE at 22 $\mu$m, finding definite excesses around 98 stars (detection rate of 3.5$\%$).

In Section 2, we describe the process we used to determine whether or not an infrared excess was present. In Section 3, we elaborate on the age-determination method we used, and explain how we compiled our target list. In Section 4, we put our results in the context of current debris disk research. In Section 5, we describe any potential issues and errors associated with our findings. 

\section{IDENTIFYING AN IR EXCESS}
Since most debris disks cannot be resolved, we depend on spectral energy distributions (SEDs) to identify IR excesses. SEDs were created with a fully automated fitting technique using theoretical models from Hauschildt et al. (1999) to predict stellar photospheric fluxes.  The SEDs were generated using available photometry from Hipparcos (Perryman et al. 1997), Tycho-2 (Hog et al. 2000), 2MASS (Cutri et al. 2003), WISE (Cutri et al. 2012), and (when available) IRAS (Helou \& Walker 1988).  Stellar radii and effective temperatures are treated as free parameters to fit the observed fluxes (B, V, J, H, and K) with a $\chi$$^{2}$ minimization method. We chose not to fit the photosphere to the W1, W2, or W3 points due to saturation limits in the WISE data and the possibility of an excess at W3. 

To characterize the IR excess (or lack thereof) in WISE, we concentrated on the W4 data (22 $\mu$m)$\footnotemark{}$$\footnotetext{All candidate debris disks presented in this work show an excess at 22$\mu$m but not at 12$\mu$m}$. The WISE data release provides the W4 flux density in magnitude units. We converted these magnitudes to flux densities in Jy using published WISE zero points. Using the photospheric fluxes predicted by the $\chi$$^{2}$ fit, we defined a parameter SNR:

\begin{equation}
SNR=\frac{W4[Jy]-W4_{phot}[Jy]}{N4[Jy]}
\end{equation}

\noindent
where N4 is the noise (error) associated with each W4 measurement and W4$_{phot}$ is the predicted 22 $\mu$m photospheric value. We choose to define a candidate excess source as one for which SNR$>$5. This constitutes a 5$\sigma$ detection. These candidate excess sources were double-checked visually to make sure that the calculated excess was not due to a bad photospheric fit.$\footnotemark{}$$\footnotetext{Bad photospheric fits can occur due to stellar variability or to an overly coarse parameter spacing in the stellar photospheric models.}$ A blackbody was then fit to the apparent IR excess. When available, we used supplementary data from Spitzer and Herschel to better constrain the dust temperature and fractional IR luminosity. These data were downloaded from the NASA/IPAC Infrared Science Archive website (irsa.ipac.caltech.edu). Relevant data are found in Table 3. One SED representative of our sample is shown in Figure 1. 

Zuckerman et al. (2011) found that Hauschildt photopsheric models underpredicted the flux at 22 $\mu$m by $\sim$3$\%$. We used a subset of stars from Jenkins et al. (2011) - a sample of stars with known chromospheric activity - to test the Hauschildt models. Of the 868 stars in the Jenkins sample, we used 230 stars which had SNR between -1 and 1, and which had W2 fluxes $<$1 Jy (to avoid the saturation limit at 2 Jy). We used this sample of 230 stars to compare two different photosphere models - one from Hauschildt et al. (1999), and the second a linear fit to W1, W2, and W3\footnotemark{}\footnotetext{The Rayleigh Jeans tail of the stellar photosphere can be approximated by a linear fit if the temperature of a star is $\gtrsim$3500 K.}. In the end, we chose to define SNR using the ``corrected" photospheric value (W4$_{phot}$*1.03). 

Finally, the data products from WISE were individually inspected to make sure that there were no contaminating sources which could be mimicking an excess (such as a binary companion or a background galaxy). The FWHM of the PSFs for the different WISE bands are: W1(3.4 $\mu$m): 6.08$\arcsec$ ; W2(4.6 $\mu$m): 6.84$\arcsec$ ; W3(11 $\mu$m): 7.36$\arcsec$ ; W4(22 $\mu$m): 11.99$\arcsec$. Since we are considering excess in the W4 band, we use the FWHM of the PSF in that band as the contamination radius; if a point-like secondary source were found within 24$\arcsec$ (2$\times$FWHM$_{W4}$) of the target, we considered that star to be ``contaminated" and it was no longer considered to be a candidate excess source. We allowed more room for extended nearby sources such as background galaxies. In the case of a nearby extended source, we define a contamination radius to be from the center of the contaminating galaxy to the center of the target in question, and set that radius at 1$\arcmin$. Furthermore, there were several cases in which an excess was seen at W4 in the SED, but no star is apparent in the WISE data product. These stars were considered non-detections.$\footnotemark{}\footnotetext{Usually, these were cases of galactic cirrus confusion.}$  If none of the aforementioned issues were encountered, the data product is considered to be ``clean."

We also considered the possibility of an unseen background galaxy contaminating our W4 excess sources. Given that the WISE data products were checked for nearby sources, any remaining contaminating source would have to lie within the 6$\arcsec$ beam of the W4 band. Kennedy \& Wyatt (2012) predict 0.06 spurious sources in their WISE 468 targets (a contamination rate of $\sim$0.01$\%$). Therefore, for our 2,820 targets, we would expect 0.36 spurious detections due to contamination by unseen background galaxies. 

\section{THE STELLAR SAMPLE}
\subsection{Chromospheric Activity as an Age Indicator}
Stars with a deep convective zone (CZ) experience differential rotation which heats the CZ and ionizes the material in it. As this ionized material rotates, it enhances pre-existing weak magnetic fields. The strengthened magnetic field magnetizes material in the stellar wind as it leaves the system. This causes the outflowing material to rotate, resulting in a net loss of angular momentum. Over time, the star will spin-down which, in turn, reduces its magnetic activity. Thus as a star ages, it will spin down and its magnetic activity will weaken in a predictable way (Skumanich 1972, Barnes 2003)

The magnetic activity can be measured by looking at the collisonally-dominated Ca II H$\&$K absorption lines at $\sim$3900\AA. In these lines, the photosphere of the star is suppressed by absorption, and we can see the emission cores due to magnetic heating of the chromosphere. Mamajek \& Hillenbrand (2008) - hereafter MH08 - provide us with a cluster-calibrated relationship between chromospheric activity and age:

\begin{equation}
log(t)=-38.053-17.912(logR'_{HK})-1.6675(logR'_{HK})^{2}
\end{equation}

\noindent
where \textit{R}$^{\prime}$$_{HK}$ is a parameter measuring the strength of the Ca II H$\&$K emission core and \textit{t} is age in years. This relation is only valid (calibrated) if -5.1$<$log\textit{R}$^{\prime}$$_{HK}$$<$-4.0. Typical errors in age associated with this method are $\sim$60$\%$ (MH08). For some stars, there may be an additional uncertainty due to the long-period ($\sim$10 year) magnetic variations due to the stellar cycle. This is discussed further in Section 5.2.

In addition to an activity-age relation, MH08 suggests that an activity-rotation-age calculation can be used which, when used to calculate the ages of stars in binaries and open clusters, resulted in lower errors. Also, the activity-rotation-age relation takes the mass of the star (parameterized by the \textit{(B-V)} color) into account.This calculation requires that the chromospheric activity index is first used to calculate the Rossby number (Noyes et al. 1984), which can then be used to calculate a rotation period. This period is then fed into a cluster-calibrated rotation-age (``gyrochronology") relation:

\begin{equation}
P(B-V,t) = a[B-V_{0}-c]^{b}t^{n}
\end{equation}

\noindent
where MH08 found that \textit{a}=0.407, \textit{b}=0.325, \textit{c}=0.495, and \textit{n}=0.556 for a rotation period \textit{P} in days and age \textit{t} in Myr ((B-V)$_{0}$ is the de-reddened color). These constants were determined by fitting the rotation-age equation to clusters with known ages (up to the age of the Hyades - 625 Myr). It is valid for stars with 0.495$<$\textit{(B-V)}$_{0}$$<$1.4. The combined activity-rotation-age relation was calibrated to clusters up to 625 Myr, field binaries up to $\sim$10 Gyr, and field stars up to $\sim$15 Gyr. By comparing the calculated ages of field binaries, MH08 were able to quote an average error of 15$\%$ in \textit{t}. We present both the age calculated directly from the rotation-age relation of MH08 (Age$_{RHK}$) and the age calculated by using the Rossby number and the rotation-age relation (Age$_{ROT}$). All of our statistics were calculated using the latter. 

In total, we cataloged WISE data for 2,820 stars with known chromospheric activity ages from 4 sources (Pace 2013 (1,251 stars); Jenkins et al. 2011 (596 stars); Isaacson $\&$ Fischer 2010 (854 stars); Wright et al. 2004 (119 stars)). When we found an overlap between these four sources, we preferentially took chromospheric data from the most recent publication. 

\subsection{Pace 2013}
Pace (2013) collected a sample of 1,741 field stars with ages derived from stellar isochrones from the Geneva Copenhagen survey of the solar neighborhood. The goal was to constrain the age range in which chromospheric activity is a reliable age-determination tool. Of the 1,741 stars in the Pace (2013) sample, 1,251 stars have log(\textit{R}$^{\prime}$$_{HK}$) in the appropriate range for the chromospheric activity relation from MH08 (-5.1$<$log(\textit{R}$^{\prime}$$_{HK}$)$<$-4.0). After examining the SEDs, we found that 132 showed evidence of some excess. Of those, 42 have clean WISE data products. Of those 42 debris disks, 27 are previously unreported in the literature.

\subsection{Jenkins et al. 2011}
Jenkins et al. (2011) assembled a catalog of 890 stars with chromospheric emission measured with an echelle spectrograph. Their purpose was to calibrate their measured S-index to the Mount Wilson S-index, thus allowing them to derive a value for log(\textit{R}$^{\prime}$$_{HK}$)for their sample. Of the 890 stars in the Jenkins et al. sample, 93 are also in the Pace (2013) sample$\footnotemark{}$$\footnotetext{We compared the \textit{R}$^{\prime}$$_{HK}$ values for stars which appeared in both the Pace (2013) catalog and the Jenkins et al. (2011) catalog. We found that they agreed to within $\sim$1$\%$.}$. Of the 797 remaining stars, 596 have \textit{R}$^{\prime}$$_{HK}$ values in the appropriate range for our chromospheric activity relation. Of those 596, 33 have excesses in W4 and clean WISE data products. Of those, 30 are presented here for the first time.

\subsection{Isaacson \& Fischer 2010}
Isaacson \& Fischer (2010) collected spectral data for over 2,600 stars. Their goal was to determine if the ``jitter" in their chromospheric activity measurements was due to the presence of a planetary system. Of the 2,647 stars in the Isaacson \& Fischer (2010) sample, 420 are also either in the Pace (2013) sample or the Jenkins et al. (2011) sample. Of the remaining 2,227 stars, 854 are on the main sequence (according to SIMBAD) and have \textit{R}$^{\prime}$$_{HK}$ values in the appropriate range. Of the 854 stars we examined, 20 have W4 excesses and clean WISE data products. Of that subsample, 14 have no previous mention in the literature.

\subsection{Wright et al. 2004}
Wright et al. (2004) constructed a sample of over 1,200 F, G, K, and M type stars with chromospheric activity measurements derived from 18,000 spectra from Lick and Keck Observatories. Of the 1,204 stars in the Wright et al. (2004) sample, 119 are unique targets with \textit{R}$^{\prime}$$_{HK}$ in the appropriate range. Three stars have evidence of an IR excess at W4 and clean data products. Of those 3 stars, all are presented here for the first time.

\section{DISCUSSION}
\subsection{Comparison to Protoplanetary Disks}
The following discussion is taken largely from the Annual Review article by Wyatt (2008): The IR excess seen in debris disks differs significantly from that seen in protoplanetary disks. The most obvious difference is that protoplanetary disks consist largely of gas and sub-micron sized dust grains. These grains radiate very efficiently (due to their high surface area-volume ratio). Thus the fractional IR luminosities ($\tau$) seen in protoplanetary disks are several orders of magnitude higher than those seen in debris disks and higher yet compared to the Sun's Kuiper Belt (KB)($\tau$$_{proto}$$\sim$10$^{-2}$, $\tau$$_{debris}$$\sim$10$^{-5}$, $\tau$$_{KB}$$\sim$10$^{-7}$). In addition to their higher IR luminosities, protoplanetary disks also have shorter lifetimes than debris disks. Most protoplanetary disks are completely dispersed $\sim$10 Myr after their formation due to mass loss and gas accretion onto the star (Wyatt 2008). Thus the circumstellar disks we see around $>$10 Myr stars must be second generation (Zuckerman 2001).

\subsection{Models of Dust Production}
The dust observed in debris disks is thought to have formed from a collisional cascade in the planetary disk. The generally accepted model of dust formation is as follows: Once the gas in the disk has been dispersed, rocky planetesimals can begin growing. At first, the disk experiences runaway growth as larger objects grow at faster rates due to gravitational focusing. Once objects reach $\sim$1000 km in size, they undergo oligarchic growth (Wyatt 2008). In this phase, a few large planetesimals clean out the small ($<$100 km) bodies in their neighborhood, and grow slowly into protoplanet-sized objects. At this stage, the large objects in the forming planetary system can dynamically stir the leftover small objects. These small objects will eventually be given enough of a velocity kick for the resulting collisions between small bodies to be destructive. The cold dust that we see around stars with ages $>$10 Myr is likely the result of these collisions.

\subsection{Models of Dust Removal}
Dust can be removed from the system in several ways. The most effective mechanism is due to radiative forces from the star. Once collisions between intermediate ($<$100km) sized bodies begins, a collisional cascade is triggered. Two bodies collide to create a population of smaller bodies, which collide with each other to create even smaller bodies, and so on until the debris reaches the blow-out radius. At this point, the dust can be ejected from the system by radiative forces. Dust can also be destroyed by spiraling toward the star as a result of Poynting-Robertson drag and ultimately being vaporized. Small particles can also be carried out by a strong stellar wind. However, since the dust collision timescale for observed debris disks is much smaller than the timescale for Poynting-Robsertson drag, most dust loss is ``collision dominated." (Zuckerman 2001).

\subsection{Observations}
We examined the WISE fluxes of 2,820 stars with chromospheric activity measurements. Of those, 98 have excesses at 22 $\mu$m (3.5$\%$). This detection rate agrees with the rate reported by Trilling et al. (2008) for 24$\mu$m excesses around 184 FGK stars (3.8$\%$)$\footnotemark{}$$\footnotetext{The Trilling et al. 2008 sample of debris disks were measured by Spitzer at 24 $\mu$m. Since Spitzer at 24 $\mu$m was $\sim$10 times more sensitive than WISE at 22 $\mu$m, we expect our detection rate - after correction for the sensitivity difference - is actually somewhat higher than Trilling's. In addition, Trilling et al. used a 3$\sigma$ detection limit to define IR-excess, while we utilize a 5$\sigma$ detection limit.}$. The ages of our debris disks range from 24 Myr to 9.1 Gyr, with an average of 2.7 Gyr. The debris disk parameters are found in Table 1, and the associated WISE data are found in Table 2. In Figure 2, we display the distribution of sample stars and debris disk stars as a function of chromospheric activity. We see the expected distribution of stars, with most of our sample being inactive (log(\textit{R}$^{\prime}$$_{HK}$)$<$-4.8) and a hint of the Vaughan-Preston gap at intermediate activity levels (Vaughan \& Preston 1980). It is clear that, relative to the background sample, debris disks are found preferentially around active stars. This is expected, since more active stars tend to also be younger. The distribution of the sample as a function of age is shown in Figure 3 and indeed, we see that debris disks are found preferentially around younger stars.  By fitting a logarithmic decline to our histogram of debris disk ages, we find that the number of debris disks declines as e$^{t0/t}$ where t$_{0}$$\sim$175 Myr (Figure 4). This is similar to the results from Su et al. 2006 who found that t$_{0}$$\sim$150 Myr.

We examined the evolution of the infrared luminosity fraction $\tau$ (=L$_{IR}$/L$_{bol}$) with age, where $\tau$ is determined from the following formula:

\begin{equation}
\tau=\frac{(\nu F_{\nu})_{peak, dust}}{(\nu F_{\nu})_{peak, star}}
\end{equation}

\noindent
where \textit{R}$_{star}$ and \textit{T}$_{star}$ are determined by the best-fit photospheric SED, \textit{T}$_{dust}$ is determined by the best-fit blackbody to the IR excess, and \textit{R}$_{dust}$ is determined from the equation$\footnotemark{}$$\footnotetext{This equation assumes that the dust is in a thin ring of inner radius \textit{R}$_{dust}$.}$:

\begin{equation}
R_{dust}=\frac{R_{star}}{2}(\frac{T_{star}}{T_{dust}})^2
\end{equation}

A histogram of dust radii for those systems with several mid-IR observations is shown in Figure 5. Figures 6, 7, and 8 show a visualization of the dust radii compared with the Solar System's Kuiper and Asteroid Belts. The mass of the dust can be derived given certain assumptions about the observed dust:

\begin{equation}
M_{dust}=\frac{F_{\nu}d^{2}}{\kappa_{\nu}B_{\nu}(T_{d})}
\end{equation}

\noindent
where \textit{d} is the distance from Earth and $\kappa$ is the dust opacity, and (in the literature)is often assumed to be 1.7 cm$^{2}$g$^{-1}$ at 800 $\mu$m (Zuckerman \& Becklin 1993). At shorter wavelengths, the opacity rises to $\sim$5 cm$^{2}$g$^{-1}$ for a 200 K body (Pollack et al. 1994). The opacity curves from Pollack et al. (1994) employ assumptions about grain size and composition of the dust. However Rhee et al. (2007) provides a relation between the dust mass and the fractional IR luminosity $\tau$ (=L$_{IR}$/L$_{bol}$):

\begin{equation}
\frac{\tau}{M_{dust}}\propto\frac{1}{\rho aR_{dust}^{2}}
\end{equation}

\noindent
where \textit{a} is the characteristic radius of the grains. This formula was confirmed empirically using stars with dust masses measured directly from the submillimeter flux, assuming that the dust is in an optically thin ring and that the radius and density of the grains do not change significantly from star to star (see their Figure 4). We similarly calculate dust masses for those stars in our sample with well defined values of $\tau$ (those stars whose IR excess could be fit with a unique blackbody curve$\footnotemark{}$$\footnotetext{Many sources had single-channel excesses and could not be fit with a unique blackbody. Further mid-IR observations are needed to constrain the properties of these sources (see Section 5.2)}$). These masses are listed in Table 3.

We expect the amount of dust (parameterized by the fractional infrared luminosity) to decline as a function of stellar age. Over time, there should be more collisions between planetestimals, leading to more debris being ground down to the blow-out size and ejected from the system by radiative forces. This decline in $\tau$ with age has been seen in A stars and solar-type stars (e.g. Su et al. 2006, Bryden et al. 2006, Rhee et al. 2007). We are able to reproduce this trend with our data; the dustiest debris disks are found around young stars ($<$2 Gyr). 

\section{Stellar Characteristics}
\subsection{Spectral Type}
A summary of our findings broken down by spectral type can be found in Table 4. We find that debris disks were detected predominantly around K-type stars with $\sim$2.5$\sigma$ significance. This seems like a strange result. Since a typical WISE sensitivity limit is 6mJy, a detectable excess would have to be at least 6mJy above the photosphere. For stars at the same distance from Earth, this excess would constitute a higher percentage of the photosphere (a higher $\tau$) for a dim K star than for a bright F star. In other words, a higher $\tau$ would be required to constitute a significant excess around a dim K star, while a lower $\tau$ would be necessary to be detectable at that same level around an F-type star. Thus we are sensitive to smaller $\tau$ in F stars than in K stars. Since our sensitivity is set by the flux of the photosphere at 22 $\mu$m (for a given distance), we should be able to quantify our sensitivity bias. The flux of the photosphere is defined by:

\begin{equation}
F_{\nu}=\pi B_{\nu}(\frac{R}{d})^{2}
\end{equation}

\noindent
where \textit{R} is the radius of the star and \textit{d} is the distance to Earth. Since our distance distribution shows no strong dependence on spectral type, we can assume that our F stars and K stars are essentially at the same distance. We also assume that all stars in Table 1 have reached the Rayleigh Jeans tail by 22 $\mu$m so that B$_{\nu}$$\propto$T. Thus:

\begin{equation}
\frac{F_{\nu, F}}{F_{\nu, K}}=\frac{T_{F}}{T_{K}}(\frac{R_{F}}{R_{K}})^{2}
\end{equation}

Using average values for our F and K stars (\textit{R}$_{F}$=1.05R$_{\sun}$, \textit{R}$_{K}$=0.84R$_{\sun}$, \textit{T}$_{F}$=6600K, \textit{T}$_{K}$=4100K), we find that \textit{F}$_{\nu,F}$/\textit{F}$_{\nu,K}$ = 2.5. Thus we are 2.5 times more sensitive to F stars than to K stars; Any discrepancy in the detection rate between F stars and K stars would only be magnified when this sensitivity is taken into account. Trilling et al. (2008) also found that K stars had a higher debris disk detection rate than G  stars (albeit with a smaller sample size). 

\subsection{Distance from Earth}
If young stars are preferentially found closer to Earth, then we would be more sensitive to younger stars, and thus our sample would be biased. We examine this possibility in Figure 9. In Figure 9, we can see that there is no obvious correlation between distance and age$\footnotemark{}$$\footnotetext{All distances were calculated from Hipparcos parallaxes (Perryman et al. 1997).}$. Thus we do not believe that a parallax effect is biasing our data. 

\subsection{Metallicity}
Since debris disks are found around the dustiest stars, it might be expected that they would also be preferentially found around stars with high metallicities. We gathered metallicity data for as many stars in our sample of 2,820 as possible ($\sim$2,000 stars) from Anderson et al. (2012), and examined the dependence of debris disk incidence on metallicity; no correlation is apparent (Figure 10). This result agrees with the findings of Greaves et al. (2005), who found no correlation between debris disk incidence and metallicity, even when there was a correlation between giant planet incidence and stellar metallicity. 

\subsection{Planet Hosts}
Anderson et al. (2012) provide data on known planet hosts; 136 of our 2,820 stars ($\sim$5$\%$) are known  hosts to one or more substellar objects. Four stars out of 98 debris disks presented here ($\sim$4$\%$) are planet hosts, and all have only one known giant planet. This is a surprising result, since we believe that debris disks are signposts of planet formation. The existence of planets, especially giant planets, helps to dynamically stir the population of small bodies to the velocity necessary for a collisional cascade. One therefore might expect to see a high percentage of planet hosts in our debris disk sample. This non-correlation between planet hosts and debris disk hosts was also noted by Bryden et al. (2009). 

HIP 3391 is host to a 1.56M$_{J}$ planet with a semi-major axis of 1.28 AU (Tamuz et al., 2008). HIP 3391 also has a 22 $\mu$m excess, but without additional data points we can only provide an upper limit on the temperature (and a lower limit on the semi major axis of the dust). For a maximum dust temperature of 200 K, the inner semi major axis of the dust is at 2.1 AU. We can thus say that the dust is likely outside the orbit of the planet. We then refitted the dust blackbody, and found a minimum temperature of 45 K, corresponding to a maximum inner semi major axis of the dust of 42 AU.

HIP 7978 is orbited by a 0.9M$_{J}$ planet at 2.022 AU (Butler et al., 2011)and a 55 K belt of dust 30 AU from the host star (this work). Additional data were taken from the IRS, IRAS and MIPs catalogs and corroborate the IR excess seen by WISE. Thus we are confident about the quoted dust temperature and radius.

HIP 90593 hosts a 0.67M$_{J}$ planet orbiting 2.24 AU from its host star (Fischer et al. 2009). It also has a single-channel excess indicating a debris disk with temperature 47 K$<$\textit{T}$<$200 K, corresponding to a range of inner radius 2.62 AU$<$\textit{R}$_{dust}$$<$47.4 AU. The dust is likely located outside of the planet.

HIP 118319 is host to a 0.71M$_{J}$ which orbits its host star at 0.233 AU (Johnson et al. 2010). It also has a 137 K debris disk (confirmed by a Spitzer 70 $\mu$m data point presented by Bryden et al. 2009) at a minimum radius of 8.43 AU, well outside the orbit of the planet.

The results are consistent with a dust location analogous to the Kuiper belt rather than the asteroid belt. Thus our results seem consistent with the physical picture of a debris disk (though we are working with small-number statistics and assuming that there are no undetected close-in planets). A visualization of the debris disks found around planet hosts with ill-determined dust locations can be seen in Figure 11. Note that the dust radii in the top two panels of Figure 11 represent the minimum radii of the dust orbits (or the maximum temperature for the dust blackbody fits).

\subsection{Lithium Abundances}
The abundance of Li absorption at 6707.8 $\AA$ has been cited as a possible tracer of stellar age (e.g. Zuckerman \& Song 2004). Twenty-eight stars in our debris disk sample have lithium abundances available in the literature. We were also able to measure lithium abundances for 8 stars in our debris disk sample using the Hamilton Echelle Spectograph on the 3 m Shane Telescope at Lick Observatory. We reached typical sensitivities of 10-15 m$\AA$. The relevant data are listed in Table 5. We compared the Li abundance with the age determined from the activity-rotation-age relation of MH08, and see that the Li abundance decreases as a function of age (see Figure 12). Four of our stars which appear to be old according to their low levels of chromospheric activity also have no detectable lithium. This correlation supports our use of chromospheric activity as an age indicator.

\section{ISSUES AND WARNINGS}

\subsection{Chromospheric Activity Variations}
We know that our Sun undergoes variations in magnetic activity over a $\sim$ 11 year period. It is assumed (and in some cases, observed) that other solar-type stars experience similar short-term variations (Duncan et al. 1991). If snapshots of stellar activity are taken, there is no way to tell if the star is being observed during an active period, or a quiet period. Only by taking long ($\sim$10 year) baseline observations can we be certain that we are measuring the average magnetic activity of that star. Few such surveys have been conducted (e.g. Wright et al. 2004), and future surveys will be limited by telescope availability (a spectral resolution of at least 1 $\AA$ is needed to observe the Ca II H \& K emission cores which indicate magnetic activity) and the willingness of observers to spend decades on a single sample of stars.  This systematic uncertainty introduces an extra source of error in the calculation of age from chromospheric activity. For the Sun, this variation amounts to a 20$\%$ uncertainty in age as calculated from the chromospheric activity during quiescence and during a period of high activity (see Vican 2012 for further discussion of this issue).

\subsection{Issues with SED Fitting}
For single band excesses, many different blackbodies could theoretically be fit to the same data point (see Figure 13). The excess we see could be from warm dust emission, or the Wein tail of a blackbody for cooler dust. In some cases, we were able to add data from previous studies (Spitzer, IRAS, etc.) to constrain a blackbody fit to the dust. For those stars for which we found only one data point in excess (i.e. W4), we fit the maximum temperature blackbody, which corresponds to a lower limit to $\tau$ (= L$_{IR}$/L$_{bol}$). For these stars, it is difficult to say how relevant our calculations for \textit{R}$_{dust}$ and \textit{M}$_{dust}$ may be, but we include these calculations in Table 1. Follow-up observations with mid-IR instruments such as those on SOFIA might help resolve these discrepancies by providing a second data point with which one could constrain a blackbody dust fit. Physical maps of the dust from ALMA would be the best way to determine the dust radius, since the blackbody radius often underestimates the true radius of the dust (Rodriguez \& Zuckerman 2012).

\section{CONCLUSIONS AND FUTURE WORK}
We examined 2,820 solar type (F, G, and K type) stars using WISE to search for an infrared excess at 22 $\mu$m. We found 98 stars with a clear WISE excess at 22 $\mu$m (a detection rate of 3.5$\%$), 74 of which are presented here for the first time. 

An IR excess at 22 $\mu$m is indicative of either hot dust or the Wein tail of a cold dust component. For debris disks with only the 22 $\mu$m data point in excess, follow-up observations are necessary to constrain the properties of the dust. If the dust is truly hot ($\sim$200 K), the drop-off in flux at long wavelengths will make detections with sub-mm instruments (e.g. ALMA, SCUBA-2) unlikely (although a non-detection could also help to constrain the dust temperature; see Bulger et al. 2013). Mid-IR instruments such as FORCAST on SOFIA could be used to confirm the presence of cold dust. It is important to understand the temperature evolution of the dust, especially for those stars which seem to be old ($>$2 Gyr). These old systems are either (1) the tail end of the steady-state evolution of solar type debris disks or (2) the result of cataclysmic collisions between two large rocky bodies (e.g. BD +20 307, Song et al. 2005). Observations in the mid-IR and sub-mm to determine the dust temperature will help distinguish between formation mechanisms.

Partial support for this work was supported by a NASA grant to UCLA and by an NSF Graduate Research Fellowship to Laura Vican. This project made use of the SIMBAD and Two-Micron All Sky Survey (2MASS) databases and VIZIER search engine, operated by CDS in France. The Wide-Field Infrared Survey Explorer is a joint project between the University of California, Los Angeles, and the Jet Propulsion Laboratory, funded by NASA. 

We thank Ben Zuckerman for his valuable insight.

\clearpage

\begin{landscape}
\begin{deluxetable}{cccccccccccccccccccc}
\tabletypesize{\scriptsize}
\tablewidth{0pt}
\tablecaption{Chromospheric Activity Targets with IR Excesses}
\tablehead{
\colhead{HIP}&
\colhead{B-V}&
\colhead{SpT}&
\colhead{log\textit{R}$^{\prime}$$_{HK}$}&
\colhead{Ref}&
\colhead{Age$_{RHK}$}&
\colhead{Age$_{ROT}$}&
\colhead{P$_{calc}$}&
\colhead{T$_{star}$}&
\colhead{R$_{star}$}&
\colhead{L$_{IR}$/L$_{bol}$}&
\colhead{T$_{dust}$}&
\colhead{R$_{dust}$}&
\colhead{d}&
\colhead{Disk ref}\\
\colhead{}&
\colhead{}&
\colhead{}&
\colhead{(dex)}&
\colhead{}&
\colhead{(Myr)}&
\colhead{(Myr)}&
\colhead{(days)}&
\colhead{K}&
\colhead{R$_{\sun}$}&
\colhead{E-05}&
\colhead{K}&
\colhead{AU}&
\colhead{pc}&
\colhead{}}
\startdata
296&7.60E-01&G8V&-4.5&P13&608&1023&13.32&5700&0.85&5.2$\tablenotemark{a}$&200$\tablenotemark{a}$&1.61$\tablenotemark{a}$&40&none\\
544&7.50E-01&K0V&-4.384&I10&266&337&7.02&5700&0.82&11.2&200&1.55&14&T08\\
682&6.30E-01&G2V&-4.359&P13&219&147&3.57&6000&1.03&14.7&100&8.62&40&C09\\
1365&7.87E-01&G5&-5.029&I10&7144&7931&43.8&5400&3.93&9.6$\tablenotemark{a}$&200$\tablenotemark{a}$&6.66$\tablenotemark{a}$&137&none\\
1481&5.40E-01&F8&-4.36&P13&221&98&1.98&6200&1.05&9.8&200&2.35&41&Z11\\
3391&7.30E-01&G5V&-5.021&P13&6994&6855&37.57&5800&1.09&5.9$\tablenotemark{a}$&200$\tablenotemark{a}$&2.13$\tablenotemark{a}$&44&none\\
5227&8.56E-01&G5V&-4.016&I10&10&24&1.76&5000&4.61&13.1$\tablenotemark{a}$&200$\tablenotemark{a}$&6.70$\tablenotemark{a}$&132&none\\
5373&8.50E-01&K0V&-4.311&P13&150&211&5.99&5400&0.75&8.4$\tablenotemark{a}$&200$\tablenotemark{a}$&1.27$\tablenotemark{a}$&35&none\\
5740&6.03E-01&G3V&-5.01&J11&6787&3824&20.96&6000&0.86&8.2$\tablenotemark{a}$&200$\tablenotemark{a}$&1.80$\tablenotemark{a}$&69&none\\
5881&6.71E-01&G5&-4.8&W04&3202&3051&21.63&5800&0.77&8.7$\tablenotemark{a}$&200$\tablenotemark{a}$&1.51$\tablenotemark{a}$&59&none\\
6276&7.50E-01&G0&-4.284&I10&120&163&4.65&5600&0.75&7.2&200&1.37&35&Z11\\
6795&7.80E-01&K0V&-4.507&I10&637&1117&14.34&5500&0.72&12.2$\tablenotemark{a}$&200$\tablenotemark{a}$&1.27$\tablenotemark{a}$&41&none\\
6856&9.10E-01&K1V&-4.324&P13&166&224&6.52&4900&0.76&12.1$\tablenotemark{a}$&200$\tablenotemark{a}$&1.06$\tablenotemark{a}$&37&Z11*\\
7576&7.97E-01&G5&-4.41&J11&323&502&9.29&5500&0.77&7.5&200&1.35&24&P09\\
7978&5.30E-01&F8V&-4.731&P13&2323&1285&7.82&6400&0.99&30.9&55&31.17&17&R07\\
8867&1.01E+00&G5&-4.654&P13&1556&2568&27.84&4900&0.69&9.4$\tablenotemark{a}$&200$\tablenotemark{a}$&0.96$\tablenotemark{a}$&22&none\\
8920&5.10E-01&G0&-4.471&I10&499&394&3.03&6100&1.26&400&&&92&O12\\
9141&6.60E-01&G3&-4.202&P13&59&78&2.66&5900&0.91&10&145&3.5&42&Z11\\
10977&9.20E-01&K2&-4.821&P13&3505&4913&37.75&5200&0.73&10.4$\tablenotemark{a}$&200$\tablenotemark{a}$&1.15$\tablenotemark{a}$&31&none\\
12198&6.20E-01&G5&-4.948&I10&5632&3612&21.28&5900&1.17&7.8$\tablenotemark{a}$&200$\tablenotemark{a}$&2.37$\tablenotemark{a}$&75&none\\
14684&8.10E-01&G0&-4.4&J11&300&456&8.92&5600&0.77&9.4&140&2.86&40&Z11\\
14809&7.10E-01&G5&-4.377&I10&252&273&5.9&5900&0.97&6.9&200&1.96&49&Z11\\
17439&8.80E-01&K1V&-4.496&P13&590&1121&15.85&5300&0.77&19.4&45&24.83&16&none\\
17903&8.17E-01&G8V&-4.41&J11&323&514&9.62&5600&0.82&8.6$\tablenotemark{a}$&200$\tablenotemark{a}$&1.49$\tablenotemark{a}$&46&none\\
18828&0.86&K0V&-4.67&I10&1697&2863&26.47&5400&0.7&9.7$\tablenotemark{a}$&200$\tablenotemark{a}$&1.19$\tablenotemark{a}$&43&none\\
19793&0.657&G3V&-4.432&I10&379&412&6.78&5900&1.04&11$\tablenotemark{a}$&195$\tablenotemark{a}$&2.21$\tablenotemark{a}$&46&none\\
20737&0.85&K0V&-4.285&P13&121&188&5.62&5400&0.77&20.1$\tablenotemark{a}$&200$\tablenotemark{a}$&1.31$\tablenotemark{a}$&39&none\\
21091&0.665&G0&-4.43&I10&374&419&6.96&6000&0.94&11.4$\tablenotemark{a}$&200$\tablenotemark{a}$&1.97$\tablenotemark{a}$&67&none\\
22320&0.82&G8&-5.042&P13&7378&8509&47.19&5600&0.89&7.4$\tablenotemark{a}$&200$\tablenotemark{a}$&1.62$\tablenotemark{a}$&55&none\\
22787&0.79&K0V&-4.72&P13&2200&3275&26.64&5600&0.78&6.7&200&1.42&26&C09\\
23243&0.683&G3&-4.96&J11&5853&5031&29.33&5800&0.94&7.9$\tablenotemark{a}$&200$\tablenotemark{a}$&1.84$\tablenotemark{a}$&70&none\\
26990&0.59&G0V&-4.233&P13&78&64&1.99&6100&1.08&10$\tablenotemark{a}$&200$\tablenotemark{a}$&2.34$\tablenotemark{a}$&52&none\\
27134&0.849&G5&-4.09&J11&21&55&2.81&5200&0.91&13$\tablenotemark{a}$&200$\tablenotemark{a}$&1.43$\tablenotemark{a}$&50&none\\
27429&0.554&G0&-4.758&I10&2645&1471&10.02&6300&1.12&8.1$\tablenotemark{a}$&200$\tablenotemark{a}$&2.58$\tablenotemark{a}$&94&none\\
29391&0.617&G1V&-4.48&J11&531&506&6.95&6100&1.01&7.3$\tablenotemark{a}$&200$\tablenotemark{a}$&2.18$\tablenotemark{a}$&73&none\\
29442&0.836&K0V&-5.06&J11&7722&8982&49.43&5500&1.05&10.1$\tablenotemark{a}$&200$\tablenotemark{a}$&1.85$\tablenotemark{a}$&76&none\\
29754&0.618&G2&-5.02&J11&6975&4247&23.21&6000&1.3&8.8$\tablenotemark{a}$&200$\tablenotemark{a}$&2.72$\tablenotemark{a}$&99&none\\
30030&0.587&G0&-4.177&I10&47&46&1.63&6100&1.05&5.1&200&2.27&50&Z11\\
33690&0.806&K0IV-V&-4.498&P13&598&1087&14.52&5600&0.82&20.6&94&6.77&18&R07\\
34147&0.705&G3V&-5.07&J11&7905&7051&36.8&6000&1.17&9$\tablenotemark{a}$&200$\tablenotemark{a}$&2.45$\tablenotemark{a}$&92&none\\
36129&0.845&G5&-5.003&I10&6656&7943&46.5&5500&0.81&9.7$\tablenotemark{a}$&200$\tablenotemark{a}$&1.42$\tablenotemark{a}$&41&none\\
36312&0.54&F7V&-4.46&J11&463&301&3.74&6300&1.11&13.9$\tablenotemark{a}$&200$\tablenotemark{a}$&2.56$\tablenotemark{a}$&90&none\\
36515&0.639&G3V&-4.385&P13&267&223&4.61&5900&0.91&4&200&1.84&22&P09\\
36948&0.74&G3&-4.317&P13&157&183&4.9&5700&0.8&271&64&14.75&35&R07\\
38072&0.628&G2V&-4.51&J11&650&665&8.34&5900&0.98&13.8$\tablenotemark{a}$&200$\tablenotemark{a}$&1.98$\tablenotemark{a}$&81&none\\
41351&0.851&G5&-4.59&J11&1077&1968&21.24&5300&0.7&10.2$\tablenotemark{a}$&200$\tablenotemark{a}$&1.14$\tablenotemark{a}$&43&none\\
43371&0.772&G5&-4.47&J11&496&839&12.08&5700&0.89&37.6$\tablenotemark{a}$&200$\tablenotemark{a}$&1.68$\tablenotemark{a}$&66&none\\
44279&0.77&G8IV&-4.91&J11&4951&5745&35.79&5700&1&15.2&200&1.89&65&none\\
45621&0.869&K0&-4.45&W04&431&782&12.81&5400&1.09&7.8$\tablenotemark{a}$&200$\tablenotemark{a}$&1.85$\tablenotemark{a}$&33&none\\
45749&0.83&K0V&-4.755&P13&2608&3892&30.62&5300&0.84&10.3$\tablenotemark{a}$&200$\tablenotemark{a}$&1.37$\tablenotemark{a}$&35&none\\
45950&0.581&G0&-4.544&I10&810&639&7.07&6200&1.15&6.7$\tablenotemark{a}$&200$\tablenotemark{a}$&2.57$\tablenotemark{a}$&71&none\\
47135&0.588&G2V&-4.34&J11&189&103&2.59&6100&1.09&7&150&4.19&63&S06\\
47990&0.663&K0&-4.64&J11&1439&1609&14.83&5900&1.15&10$\tablenotemark{a}$&200$\tablenotemark{a}$&2.33$\tablenotemark{a}$&66&none\\
48133&0.894&K1V&-4.91&J11&4951&6368&42.83&5200&0.87&10.4&200&1.37&26&none\\
48423&0.72&G5&-4.468&P13&489&724&10.38&5800&0.89&11.1&150&3.09&32&C09\\
50701&0.623&G3V&-4.97&J11&6038&3876&22.32&5800&1.06&10.1$\tablenotemark{a}$&200$\tablenotemark{a}$&2.07$\tablenotemark{a}$&67&none\\
50757&0.649&F7V&-4.669&I10&1688&1708&14.92&6000&7.92&16$\tablenotemark{a}$&200$\tablenotemark{a}$&16.57$\tablenotemark{a}$&283&none\\
51783&0.713&G5&-5.06&J11&7722&7092&37.37&5700&0.87&9.8$\tablenotemark{a}$&200$\tablenotemark{a}$&1.64$\tablenotemark{a}$&62&none\\
51884&0.855&K0&-4.47&W04&496&920&13.86&5300&1.02&7$\tablenotemark{a}$&200$\tablenotemark{a}$&1.67$\tablenotemark{a}$&34&none\\
52462&0.877&K1V&-4.338&P13&186&237&6.56&5300&0.73&96.4&40&29.8&22&R07\\
52933&0.892&K0&-5.06&J11&7722&9136&52.44&5400&0.85&12.7$\tablenotemark{a}$&200$\tablenotemark{a}$&1.44$\tablenotemark{a}$&52&none\\
54459&1.09&K0&-5.054&P13&7602&8292&56.64&4600&0.58&13.8$\tablenotemark{a}$&190$\tablenotemark{a}$&0.79$\tablenotemark{a}$&26&none\\
59259&1.06&K2V&-4.299&P13&135&188&6.53&5100&1.1&15.9$\tablenotemark{a}$&200$\tablenotemark{a}$&1.66$\tablenotemark{a}$&51&none\\
59315&0.7&G5V&-4.294&P13&130&146&4.08&5800&0.83&7.3$\tablenotemark{a}$&200$\tablenotemark{a}$&1.62$\tablenotemark{a}$&38&none\\
60074&0.6&G2V&-4.385&P13&268&182&3.71&6000&0.94&107&55&26.01&29&R07\\
66676&0.588&G0V&-4.78&J11&2929&1868&13.31&6000&1.09&12.6$\tablenotemark{a}$&200$\tablenotemark{a}$&2.28$\tablenotemark{a}$&59&none\\
66765&0.86&K1V&-4.408&P13&317&515&10.03&5400&0.73&8.7&200&1.24&16&T08\\
67055&0.876&K1V&-4.56&J11&896&1686&19.9&5400&0.72&13.1$\tablenotemark{a}$&200$\tablenotemark{a}$&1.22$\tablenotemark{a}$&45&none\\
69129&0.875&G5&-4.99&J11&6412&7784&47.22&5300&0.83&10.5$\tablenotemark{a}$&200$\tablenotemark{a}$&1.36$\tablenotemark{a}$&45&none\\
71640&0.652&G3V&-4.71&J11&2092&2042&16.61&6000&1.05&7.1$\tablenotemark{a}$&200$\tablenotemark{a}$&2.20$\tablenotemark{a}$&72&none\\
73061&0.783&G8V&-4.98&J11&6224&7027&40.71&5500&2.01&6.22$\tablenotemark{a}$&200$\tablenotemark{a}$&3.53$\tablenotemark{a}$&91&none\\
73869&0.75&G5&-4.948&P13&5632&6068&36.01&5700&0.91&4.6$\tablenotemark{a}$?&200$\tablenotemark{a}$&1.72$\tablenotemark{a}$&44&H08\\
75266&1&K3V&-5.014&P13&6853&7913&52.29&5100&0.77&10.1$\tablenotemark{a}$&200$\tablenotemark{a}$&1.16$\tablenotemark{a}$&25&none\\
76280&0.669&G5&-4.904&P13&4846&4102&25.48&5800&1&9.1$\tablenotemark{a}$&200$\tablenotemark{a}$&1.96$\tablenotemark{a}$&41&none\\
76704&0.771&G5&-4.578&I10&1001&1684&17.9&5600&2.71&8.9$\tablenotemark{a}$&200$\tablenotemark{a}$&4.94$\tablenotemark{a}$&113&none\\
76757&0.605&G5&-4.617&I10&1262&1075&10.28&6100&1.08&8.1$\tablenotemark{a}$&200$\tablenotemark{a}$&2.34$\tablenotemark{a}$&73&none\\
77199&0.97&K2V&-4.155&P13&38&90&4.08&4600&1.03&16.5$\tablenotemark{a}$&200$\tablenotemark{a}$&1.27$\tablenotemark{a}$&41&none\\
77603&0.7&G2IV/V&-5.059&P13&7703&6762&35.66&5900&2.46&7.1$\tablenotemark{a}$&200$\tablenotemark{a}$&4.98$\tablenotemark{a}$&107&none\\
78466&0.63&G3V&-4.874&P13&4337&3126&20.11&5800&1.11&9.3$\tablenotemark{a}$&200$\tablenotemark{a}$&2.17$\tablenotemark{a}$&46&none\\
80129&0.795&G6IV&-5&J11&6599&7509&42.84&5600&2.01&7.9$\tablenotemark{a}$&200$\tablenotemark{a}$&3.66$\tablenotemark{a}$&92&none\\
87091&0.99&K2IV/V&-4.548&P13&831&1521&20.44&5000&0.68&13.8$\tablenotemark{a}$&200$\tablenotemark{a}$&0.99$\tablenotemark{a}$&31&none\\
87116&0.71&G6&-4.935&P13&5396&5225&31.3&5700&0.94&8.5$\tablenotemark{a}$&200$\tablenotemark{a}$&1.78$\tablenotemark{a}$&27&none\\
90593&0.68&G5&-5.017&P13&6919&5694&31.29&6000&1.25&6.4$\tablenotemark{a}$&200$\tablenotemark{a}$&2.62$\tablenotemark{a}$&65&none\\
92304&0.737&G8V&-4.45&J11&431&655&10.05&5700&1.15&9.34$\tablenotemark{a}$&200$\tablenotemark{a}$&2.17$\tablenotemark{a}$&76&none\\
96635&0.872&G8IV&-4.18&J11&48&109&4.2&5200&0.67&12.3$\tablenotemark{a}$&200$\tablenotemark{a}$&1.05$\tablenotemark{a}$&35&none\\
96854&0.7&G6IV/V&-4.961&P13&5862&5380&31.33&5800&0.99&7.1$\tablenotemark{a}$&200$\tablenotemark{a}$&1.94$\tablenotemark{a}$&42&none\\
98621&0.68&G5V&-4.923&P13&5172&4517&27.45&5800&0.84&7.3$\tablenotemark{a}$&200$\tablenotemark{a}$&1.64$\tablenotemark{a}$&38&none\\
100942&0.68&G2&-4.947&P13&5614&4810&28.44&6000&1.01&7.8$\tablenotemark{a}$&200$\tablenotemark{a}$&2.11$\tablenotemark{a}$&83&none\\
101726&0.66&G3V&-4.452&P13&436&499&7.61&5900&0.82&7.1$\tablenotemark{a}$&200$\tablenotemark{a}$&1.66$\tablenotemark{a}$&37&none\\
105388&0.65&G5V&-4.144&P13&35&51&2.06&5700&0.79&24.3&100&5.97&46&Z11\\
107457&0.727&G5&-4.46&I10&463&693&10.23&5900&0.87&8.6$\tablenotemark{a}$&200$\tablenotemark{a}$&1.76$\tablenotemark{a}$&38&none\\
113010&0.875&K1&-4.99&J11&6412&7784&47.22&5100&0.81&17.8$\tablenotemark{a}$&200$\tablenotemark{a}$&1.22$\tablenotemark{a}$&47&none\\
115527&0.71&G5&-4.394&P13&286&338&6.65&5800&0.84&7.2$\tablenotemark{a}$&200$\tablenotemark{a}$&1.64$\tablenotemark{a}$&30&none\\
116376&0.702&G5V&-4.57&J11&953&1333&14.28&5700&0.96&8.1$\tablenotemark{a}$&200$\tablenotemark{a}$&1.81$\tablenotemark{a}$&71&none\\
117247&0.882&K1V&-4.9&J11&4777&6199&41.76&5200&0.77&9.33$\tablenotemark{a}$&200$\tablenotemark{a}$&1.21$\tablenotemark{a}$&36&none\\
117481&0.5&F6&-4.413&P13&330&375&2.05&6300&1.05&5.9$\tablenotemark{a}$&200$\tablenotemark{a}$&2.42$\tablenotemark{a}$&36&none\\
117702&0.794&K1V&-5.07&J11&7905&8775&46.73&5300&0.79&11.7$\tablenotemark{a}$&200$\tablenotemark{a}$&1.29$\tablenotemark{a}$&48&none\\
118319&0.639&G2V&-5.082&I10&8123&5435&28.09&6000&1.89&7.8&137&8.43&94&B09\\
\enddata
\tablecomments{\tiny P13=Pace 2013, J11=Jenkins et al. 2011, I10=Isaacson \& Fischer 2010, W04=Wright et al. 2004, Z11=Zuckerman et al. 2011, B09=Bryden et al. 2009, H08=Hillenbrand et al. 2008, T08=Trilling et al. 2008, Rhee et al. 2007, C09=Carpenter et al. 2009, Plavchan et al. 2009. $^{a}$These stars have only one data point in excess. Thus, the dust blackbody was fit for the highest possible temperature (and consequently the lowest possible tau). \\
*Zuckerman et al. 2011 did not find this star to have an IR excess.\\
Age$_{RHK}$ is the age calculated directly from the chromospheric activity, while Age$_{ROT}$ is the age calculated from the rotation period (P$_{calc}$), which is in turn\\
 calculated from the chromospheric activity. T$_{star}$ and R$_{star}$ come from the Hauschildt et al. 1999 photosphere fits to the optical data points. L$_{IR}$/L$_{bol}$, T$_{dust}$\\
  and R$_{dust}$ come from the blackbody fit to the mid-IR data points. M$_{dust}$ is calculated for those stars with a well-determined blackbody fits using the relation\\
   from Rhee et al 2007. The distance from Earth (d) is calculated from the published Hipparcos parallaxes.}
\end{deluxetable}

\begin{deluxetable}{cccccccccccccc}
\tablewidth{0pt}
\tablecaption{WISE Data for Stars with Excesses}
\tablehead{
\colhead{(1)}&
\colhead{(2)}&
\colhead{(3)}&
\colhead{(4)}&
\colhead{(5)}&
\colhead{(6)}&
\colhead{(7)}&
\colhead{(8)}&
\colhead{(9)}&
\colhead{(10)}&
\colhead{(11)}&
\colhead{(12)}&
\colhead{(13)}\\
\colhead{HIP}&
\colhead{Star Ref}&
\colhead{W1phot}&
\colhead{W1}&
\colhead{W2phot}&
\colhead{W2}&
\colhead{W3phot}&
\colhead{W3}&
\colhead{W4phot}&
\colhead{W4}&
\colhead{W4Err}&
\colhead{W4 Ex}&
\colhead{SNR}\\
\colhead{}&
\colhead{}&
\colhead{mJy}&
\colhead{mJy}&
\colhead{mJy}&
\colhead{mJy}&
\colhead{mJy}&
\colhead{mJy}&
\colhead{mJy}&
\colhead{mJy}&
\colhead{mJy}&
\colhead{mJy}&
\colhead{mJy}&
\colhead{}}
\startdata
296&P13&721.26&778.2&344.04&414.71&66.42&65.9&18.04&21.45&0.47&3.41&7.22\\
544&I10&5367.38&6156.72&2560.27&3457.17&494.26&499.93&134.28&184.78&1.7&50.49&29.7\\
682&P13&1110.77&1158.48&557.44&651.24&100.74&98.25&27.53&41.25&0.71&13.72&19.35\\
1365&I10&874.3&929.58&398.68&524.49&80.05&82.41&21.68&27.41&0.53&5.73&10.89\\
1481&P13&1040.67&1072.23&558.27&618.5&97.78&95.35&27.92&41.35&0.51&13.43&26.19\\
3391&P13&934.35&1009.92&450.45&543.18&85.82&83.89&23.33&28.07&0.48&4.74&9.81\\
5227&I10&2146.64&2423.89&958.58&1359.92&198.75&215.31&53.69&71.09&0.8&17.4&21.86\\
5373&P13&644.48&730.95&293.89&376.83&59.01&59.53&15.98&19.77&0.46&3.78&8.24\\
5740&J11&213.23&219.33&107.01&117.31&19.34&18.82&5.29&7.1&0.36&1.82&5.03\\
5881&W04&396.03&411.04&190.92&217.64&36.38&35.17&9.89&12.75&0.35&2.86&8.13\\
6276&I10&701.85&747.29&330.61&397.87&64.75&65.1&17.57&22.61&0.53&5.04&9.6\\
6795&I10&388.81&529.53&179.32&250.12&35.68&38.73&9.62&13.56&0.4&3.93&9.81\\
6856&P13&559.89&603.52&252.23&321.32&52.82&53.22&14.34&17.88&0.42&3.54&8.54\\
7576&J11&1484.76&1613.99&684.76&870.96&136.26&132.57&36.75&48.37&0.62&11.62&18.77\\
7978&P13&5290.07&6542.38&2851.54&3847.18&492.71&561.5&141.75&216.41&1.73&74.66&43.13\\
8867&P13&1182.05&1217.55&532.51&680.8&111.52&107.96&30.27&35.74&0.62&5.48&8.91\\
8920&I10&276.4&324.11&147.9&208.23&25.99&632.34&7.42&538.11&3.66&530.69&145.04\\
9141&P13&764.2&851.7&387.29&453.04&71.75&72.29&19.89&28.25&0.46&8.37&18.07\\
10977&P13&671.34&723.58&300.72&385.25&61.57&62.29&16.51&20.39&0.42&3.88&9.32\\
12198&I10&420.69&437.21&213.2&229.58&39.5&36.84&10.95&14.15&0.39&3.2&8.2\\
14684&J11&632.44&674.67&297.91&362.86&58.35&58.45&15.83&21.12&0.37&5.29&14.22\\
14809&I10&503.44&529.53&255.13&286.11&47.27&45.91&13.1&16.82&0.48&3.72&7.79\\
17439&P13&3193.48&3504.93&1437.64&1871.78&292.39&285.93&78.38&93.74&0.86&15.37&17.83\\
17903&J11&437.51&485.61&206.09&255.47&40.36&41.38&10.95&13.53&0.42&2.57&6.1\\
18828&I10&393&432.4&179.21&223.94&35.98&36.67&9.75&12.51&0.37&2.76&7.56\\
19793&I10&748.66&763.29&379.41&422.03&70.29&66.17&19.48&27.74&0.56&8.26&14.87\\
20737&P13&596.66&643.71&272.08&345.26&54.63&56.96&14.8&23.96&0.47&9.17&19.5\\
21091&I10&396.87&422.95&199.17&223.53&35.99&35.15&9.84&14.54&0.55&4.7&8.6\\
22320&P13&357.68&391.46&168.48&202.55&33&33.32&8.95&11&0.34&2.04&6.02\\
22787&P13&1427.91&1627.4&672.62&880.44&131.74&134.51&35.75&43.41&0.63&7.66&12.25\\
23243&J11&276.97&292.34&133.53&157.81&25.44&25.4&6.91&8.84&0.27&1.92&7.14\\
26990&P13&606.13&630.21&324.34&347.49&56.98&55.94&16.27&22.89&0.37&6.62&18.09\\
27134&J11&453.46&514.15&203.13&286.91&41.59&47.47&11.15&15.38&0.3&4.23&14.33\\
27429&I10&201.64&207.15&108.45&112.45&18.74&18.2&5.41&7.33&0.35&1.92&5.5\\
29391&J11&320.66&343.79&171.59&188.17&30.15&29.88&8.61&11.23&0.37&2.62&7.04\\
29442&J11&277.26&288.33&127.87&151.68&25.44&25.17&6.86&9.25&0.34&2.39&6.94\\
29754&J11&256.74&266.62&128.85&140.26&23.28&22.49&6.36&8.51&0.39&2.14&5.48\\
30030&I10&734.69&838.47&393.14&434.65&69.07&69.33&19.72&24.39&0.54&4.67&8.7\\
33690&P13&2994.1&3254.83&1410.38&1787.97&276.23&276.16&74.96&110.46&1.11&35.49&32.12\\
34147&J11&259.54&269.58&130.25&142.87&23.54&23.4&6.43&8.84&0.29&2.41&8.31\\
36129&I10&611.6&707.76&282.07&366.56&56.13&58.35&15.14&19.22&0.44&4.08&9.32\\
36312&J11&286.18&304.43&153.91&165.1&26.6&27.27&7.67&12.79&0.33&5.12&15.36\\
36515&P13&2686.2&3042.82&1361.33&1607.05&252.19&247.21&69.9&86.75&0.94&16.85&17.98\\
36948&P13&769.41&819.39&367.01&453.04&70.85&73.07&19.25&43.92&0.67&24.67&36.93\\
38072&J11&258.83&272.83&131.17&144.86&24.3&24.16&6.74&10.57&0.3&3.83&12.94\\
41351&J11&408.76&443.29&184.01&229.79&37.43&37.21&10.03&12.53&0.46&2.5&5.42\\
43371&J11&229.69&244.28&109.56&130.9&21.15&23.3&5.75&13.72&0.36&7.98&22.03\\
44279&J11&361.23&402.8&172.31&209.19&33.26&33.06&9.04&11.54&0.3&2.5&8.35\\
45621&W04&1469.2&1520.46&669.96&906.66&134.52&140.59&36.44&45.03&0.61&8.6&14.05\\
45749&P13&756.03&833.08&340.35&456.39&69.22&72.76&18.55&23.87&0.37&5.32&14.3\\
45950&I10&397.23&414.85&213.1&227.48&37.32&35.95&10.66&13.79&0.41&3.14&7.58\\
47135&J11&410.75&435.6&219.8&234.5&38.62&38.38&11.03&14.29&0.31&3.26&10.38\\
47990&J11&424.11&446.57&214.93&244.87&39.82&38.5&11.04&14.77&0.37&3.74&10.21\\
48133&J11&1528.79&1657.61&684.82&966.42&140.21&150.11&37.59&46.91&0.69&9.32&13.43\\
48423&P13&1126.01&1093.17&542.85&639.35&103.42&99.47&28.11&42.72&0.65&14.61&22.51\\
50701&J11&434.93&481.6&209.68&259.74&39.95&41.35&10.86&14.46&0.45&3.6&8.1\\
50757&I10&1253.79&1341.11&629.22&719.44&113.71&115.75&31.08&51.69&0.6&20.61&34.36\\
51783&J11&260.14&286.21&124.09&150.57&23.96&24.35&6.51&8.85&0.39&2.34&6.07\\
51884&W04&1241.11&1369.63&558.72&714.36&113.64&112.27&30.46&36.88&0.66&6.41&9.66\\
52462&P13&1645.64&1740.48&740.83&970.87&150.67&151.17&40.39&49.51&0.61&9.13&14.86\\
52933&J11&427.67&467.62&195.02&244.42&39.16&39.81&10.61&14.97&0.46&4.36&9.46\\
54459&P13&553.42&633.12&250.91&320.44&52.62&50.6&14.64&18.49&0.54&3.85&7.12\\
59259&P13&607.34&730.95&271.41&388.81&55.94&63.84&15.03&21.12&0.47&6.09&13.09\\
59315&P13&753.36&820.14&363.19&440.7&69.2&70.16&18.81&24.29&0.42&5.48&13.11\\
60074&P13&1858.5&1929.89&932.69&1091.14&168.55&164.1&46.07&68.9&0.83&22.83&27.61\\
66676&J11&538.09&639.57&270.04&337.71&48.8&55.15&13.34&21.39&0.72&8.05&11.2\\
66765&P13&3127.27&3518.11&1426.05&1848.43&286.33&290.51&77.56&103.45&1.04&25.89&25.02\\
67055&J11&385.85&414.85&175.95&216.84&35.33&34.9&9.57&12.96&0.38&3.39&8.84\\
69129&J11&393.61&429.62&177.2&224.36&36.04&37.97&9.66&12.48&0.37&2.82&7.73\\
71640&J11&331.21&346.97&166.22&185.76&30.04&29.08&8.21&10.73&0.42&2.52&6.06\\
73061&J11&791.61&867.54&365.09&455.55&72.65&72.45&19.59&23.93&0.52&4.34&8.39\\
73869&P13&788.33&844.67&376.04&442.73&72.59&69.33&19.72&22.92&0.43&3.2&7.42\\
75266&P13&1151.08&1215.31&514.4&671.34&106.02&105.75&28.49&34.86&0.49&6.37&13.05\\
76280&P13&842.04&882.85&405.94&485&77.34&75.46&21.02&27.35&0.46&6.33&13.76\\
76704&I10&804.07&838.47&378.76&452.63&74.18&72.84&20.13&25.63&0.35&5.5&15.76\\
76757&I10&377.36&400.21&201.93&214.06&35.48&34.15&10.13&13.2&0.29&3.07&10.58\\
77199&P13&759.08&847.01&344.15&473.96&72.17&77.73&20.09&27.05&0.5&6.96&13.98\\
77603&P13&784.36&823.93&397.5&446.83&73.64&70.9&20.41&25&0.53&4.58&8.65\\
78466&P13&869.66&891.84&419.26&500.43&79.88&78.8&21.71&28.46&0.55&6.75&12.36\\
80129&J11&823.48&936.46&387.9&480.99&75.97&75.92&20.62&26.28&0.57&5.66&9.97\\
87091&P13&647.02&699.34&288.93&362.19&59.9&59.63&16.18&21.31&0.45&5.13&11.35\\
87116&P13&1932.13&2015.57&921.64&1107.84&177.92&164.63&48.34&61.75&1.06&13.41&12.63\\
90593&P13&631.64&667.25&316.99&358.87&57.28&56.22&15.66&19.89&0.44&4.23&9.66\\
92304&J11&409.81&541.36&195.48&259.02&37.74&40.41&10.25&13.84&0.44&3.58&8.2\\
96635&J11&492.56&518.91&220.64&286.64&45.17&46.6&12.11&15.99&0.51&3.87&7.67\\
96854&P13&789.58&878.8&380.66&470.48&72.52&72.86&19.71&24.83&0.56&5.12&9.21\\
98621&P13&766.64&856.42&369.59&444.77&70.42&69.46&19.14&23.45&0.57&4.31&7.54\\
100942&P13&305.13&323.81&153.13&171.14&27.67&27.44&7.56&10.09&0.38&2.52&6.66\\
101726&P13&732.56&823.17&371.25&450.96&68.78&70&19.06&24.16&0.55&5.1&9.25\\
105388&P13&509.64&576.35&243.1&304.33&46.93&49.31&12.75&19.67&0.43&6.92&15.99\\
107457&I10&787.7&877.18&399.19&479.67&73.95&75.28&20.5&27.03&0.34&6.53&19.5\\
113010&J11&445.83&546.37&199.23&281.41&41.06&46.25&11.04&15.69&0.41&4.65&11.24\\
115527&P13&1200.58&1319.22&578.8&698.77&110.27&108.37&29.97&38.53&0.6&8.56&14.24\\
116376&J11&308.29&314.98&147.05&168.32&28.39&27.45&7.71&10.16&0.44&2.45&5.58\\
117247&J11&603.47&647.27&270.32&356.24&55.35&57.39&14.84&17.94&0.49&3.1&6.35\\
117481&P13&1565.57&1630.4&841.98&893.46&145.49&140.62&41.98&54.76&0.81&12.78&15.78\\
117702&J11&346.93&374.19&156.18&194.87&31.77&31.94&8.52&11.36&0.38&2.85&7.56\\
118319&I10&587.43&592.5&294.81&322.51&53.28&51.75&14.56&19.49&0.51&4.93&9.72\\\\
\enddata
\tablecomments{P13=Pace 2013, J11=Jenkins et al. 2011, I10=Isaacson \& Fischer 2010, W04=Wright et al. 2004.\\
Columns 3, 5, 7, and 9 represent the predicted photospheric values at the W1, W2, W3, and W4 wavelengths. Columns 4, 6, 8, and 10 represent the measured fluxes from WISE. Column 11 is the published WISE uncertainty at W4. Column 12 is excess flux above the photosphere at W4 (W4-W4$_{phot}$). Column 13 is the signal to noise ratio of the W4 excess (discussed in Section 2). }
\end{deluxetable}
\end{landscape}

\begin{deluxetable}{ccc}
\tablewidth{0pt}
\tablecaption{Supplementary Data for Stars with WISE Excesses}
\tablehead{
\colhead{HIP}&
\colhead{M$_{dust}$ (M$_{Earth}$)}&
\colhead{Instrument}}
\startdata
544&4.38E-04&IRS, MIPS, IRAS\\
682&3.20E-03&IRS, MIPS, IRAS\\
1481&5.81E-04&IRS, MIPS, IRAS\\
6276&2.49E-04&IRS, MIPS, IRAS\\
7576&2.55E-04&IRS, MIPS, IRAS\\
7978&2.43E-02&IRS, MIPS, IRAS\\
9141&8.83E-04&IRS, MIPS\\
14684&6.78E-04&IRS, MIPS\\
17439&1.21E-02&IRS, MIPS, IRAS\\
22787&2.40E-04&IRS, MIPS, IRAS\\
30030&2.92E-04&IRS, MIPS, IRAS\\
33690&3.52E-03&IRS, MIPS, IRAS\\
36515&1.86E-04&IRS, MIPS, IRAS\\
36948&1.01E-01&IRS, MIPS, IRAS\\
44279&7.25E-04&IRAS \\
47135&7.40E-04&IRS, MIPS\\
48133&3.59E-04&MSX, IRAS\\
48423&8.65E-04&IRS, MIPS, IRAS\\
52462&7.24E-02&IRS, MIPS, IRAS\\
60074&7.02E-02&IRS, MIPS, IRAS\\
66765&2.72E-04&IRS, MIPS, IRAS\\
105388&3.66E-03&IRS, MIPS, IRAS\\
118319&2.27E-02&IRS, MIPS, IRAS\\
\enddata
\end{deluxetable}

\begin{deluxetable}{cc}
\tablewidth{0pt}
\tablecaption{Detection Fraction by Spectral Type}
\tablehead{
\colhead{Type}&
\colhead{Fraction}}
\startdata
F Stars&5/433 (1.15$\pm$0.4$\%$)\\
G Stars&65/1904 (3.4$\pm$0.4$\%$)\\
K Stars&28/482 (5.8$\pm$1.0$\%$)\\
\enddata
\end{deluxetable}

\begin{deluxetable}{ccccc}
\tablewidth{0pt}
\tablecaption{Lithium Data}
\tablehead{
\colhead{HIP}&
\colhead{B-V}&
\colhead{EW(H$\alpha$)}&
\colhead{EW(Li)}&
\colhead{Ref}\\
\colhead{}&
\colhead{}&
\colhead{$\AA$}&
\colhead{m$\AA$}&
\colhead{}}
\startdata
682&0.630&2.34&122&W07\\
1481&0.540&&130&T06\\
5227&0.856&&74&T06\\
5373&0.850&&50&T06\\
6276&0.750&&160&T06\\
6856&0.910&&168&T06\\
9141&0.660&2.52&195&W07\\
14684&0.810&1.76&178&W07\\
14809&0.710&&145&D09\\
17439&0.870&&0&T06\\
22787&0.800&2.15&11&W07\\
27134&0.850&0&40&T06\\
27429&0.554&2.342&81&this work\\
29442&0.836&1.598&$<$10&this work\\
29754&0.618&1.966&16&this work\\
30030&0.587&2.63&177&W07\\
33690&0.790&&0&T06\\
36129&0.845&1.495&$<$5&this work\\
36515&0.640&&81&T06\\
36948&0.740&2.32&176&W07\\
41351&0.851&1.477&$<$10&this work\\
47990&0.663&1.912&31&this work\\
48423&0.720&2.63&62&W07\\
52462&0.877&&138&T06\\
59259&1.060&&0&T06\\
59315&0.710&&152&T06\\
60074&0.600&2.79&123&W07\\
66765&0.860&&10&T06\\
73869&0.750&2.14&166&W07\\
76757&0.605&1.924&87&this work\\
77199&0.970&&420&T06\\
90593&0.680&1.825&$<$10&this work\\
96635&0.872&1.6&11&W07\\
101726&0.650&&60&T06\\
105388&0.720&&227&T06\\
115527&0.710&&133&T06
\enddata
\tablecomments{T06=Torres et al. 2006, W07=White et al. 2007, D09=Da Silva et al. 2009}
\end{deluxetable}

%\begin{figure}
%\center
%\title{Hauschildt Photosphere vs. Linear Photosphere}
%\includegraphics[width=175mm]{linfit_phot_compare_hist.eps}
%\caption{Comparison between the Hauschildt model photosphere (W4$_{Haus}$), the linear photosphere (W4$_{lin}$), and the measured 22 $\mu$m WISE flux (W4). We find that the Hauschildt model underestimates the W4 flux by $\sim$2.3$\%$. The linear fit underestimates the flux by $\sim$11.8$\%$. Our results agree with those of Zuckerman et al. (2011). We choose to use the Hauschildt model (+3.0$\%$) to define a ``corrected" SNR in excess stars.\label{fig1}}
%\end{figure}

\begin{figure}
\center
\includegraphics[width=175mm]{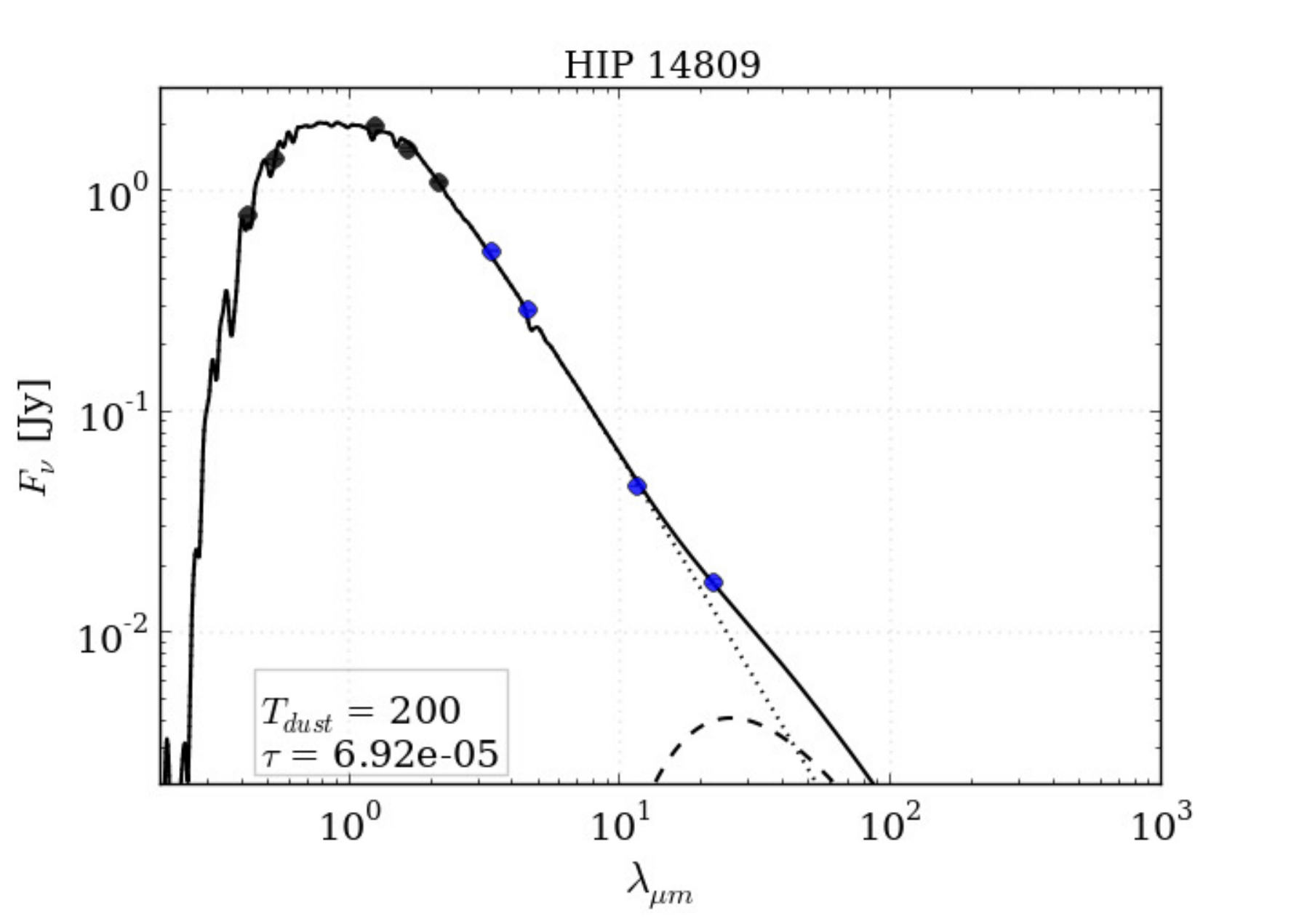}
\caption{This is an SED for HIP 14809 - one of our debris disk candidates. Black dots represent data from Hipparcos and 2MASS catalogs (at B, V, J, H, and K bands). The blue dots represent WISE data in four bands (3.4, 4.6, 11 and 22 $\mu$m).This star represents the minimum amount of dust (amount of flux above the photosphere) which we felt comfortable characterizing as a debris disk. \label{fig1}}
\end{figure}

\begin{figure}
\center
\title{Distribution of log\textit{R}$^{\prime}$$_{HK}$}
\includegraphics[width=150mm]{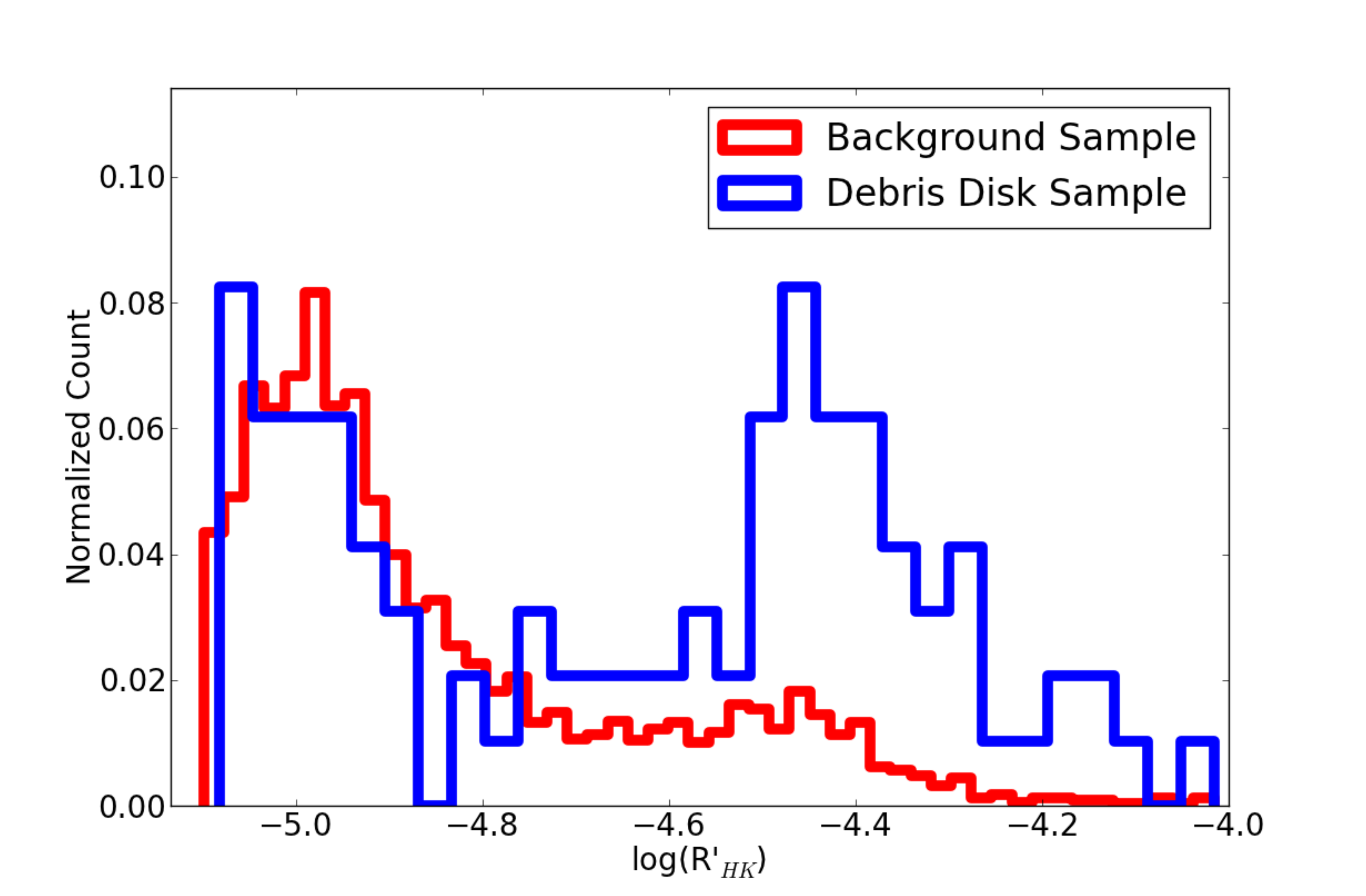}
\caption{The distribution of stars as a function of their chromospheric activity parameter. It is clear that as the background sample (which includes the debris disk stars) decreases toward higher values of logR$_{HK}$ (more active stars), the sample of debris disks does not. In fact, the distribution of debris disks appears to be somewhat bimodal.\label{fig2}}
\end{figure}

\begin{figure}
\center
\title{Distribution of Ages}
\includegraphics[width=150mm]{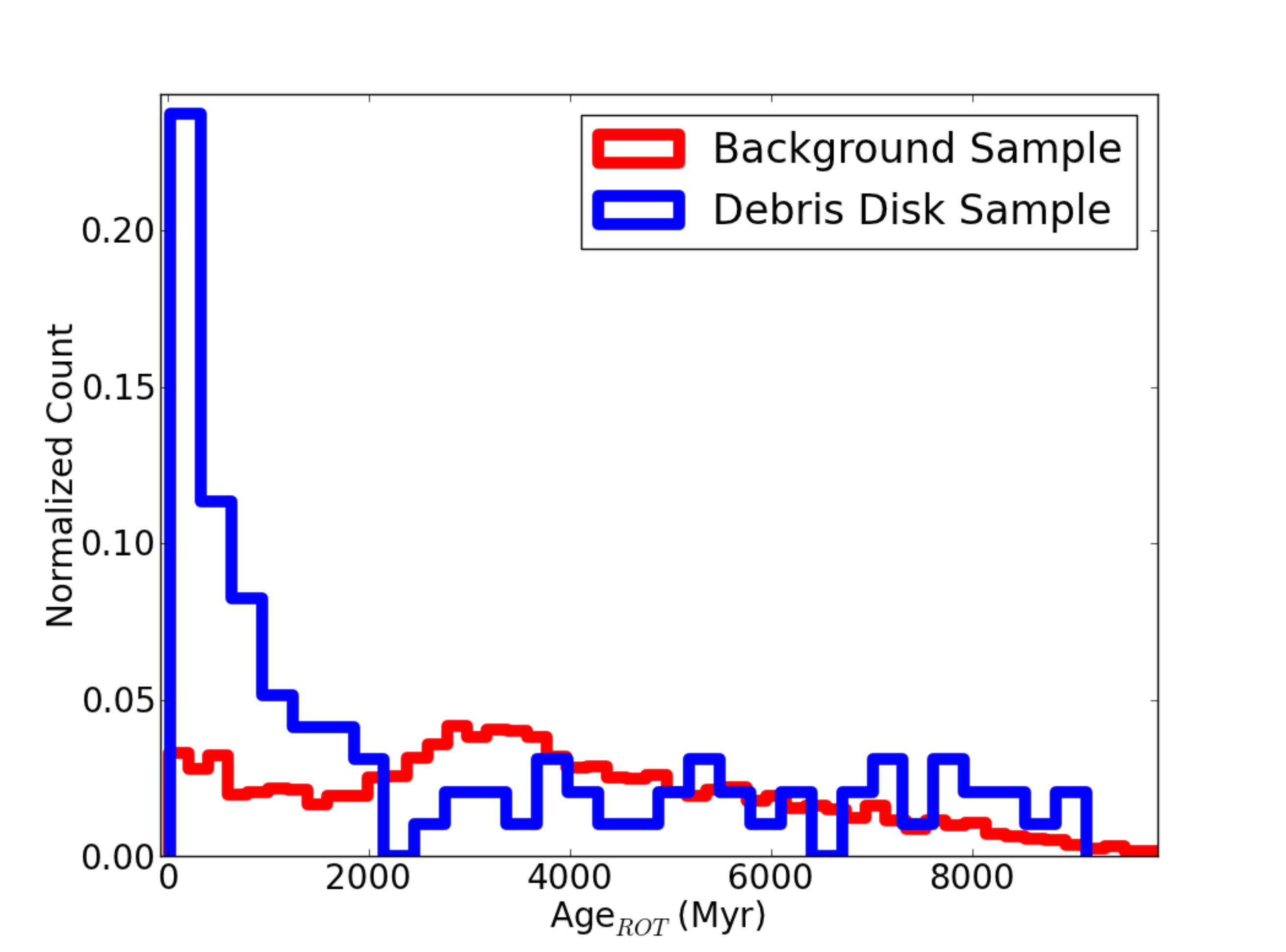}
\caption{The distribution of stars as a function of stellar age, as determined from their chromospheric activity. While the background sample (which includes debris disks stars) has a smooth distribution of ages, there is a clear peak in the distribution of debris disks at younger ages.\label{fig3}}
\end{figure}

\clearpage
\begin{figure}
\center
\title{Debris Disk Evolution}
\includegraphics[width=150mm]{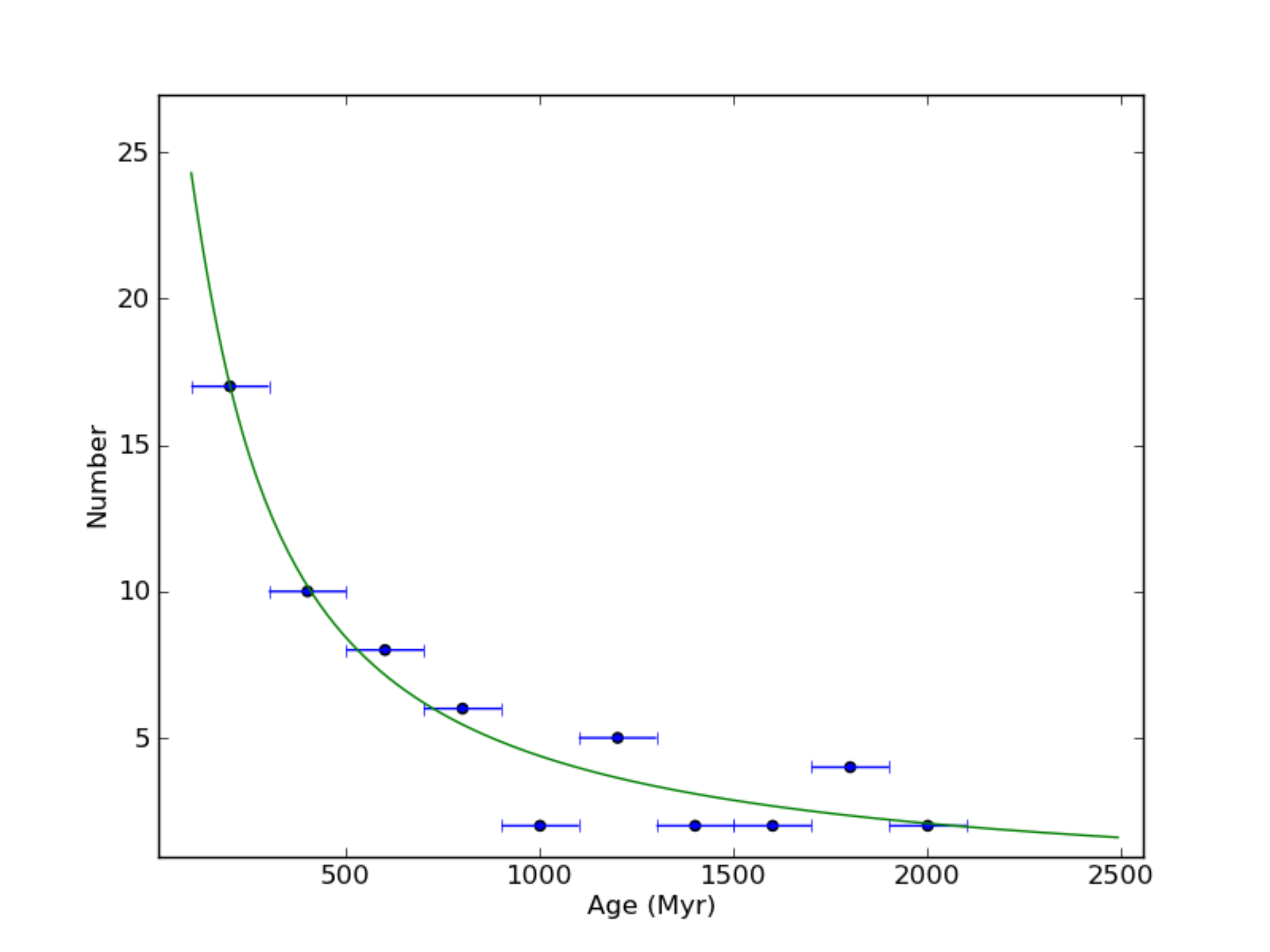}
\caption{This figure shows the evolution of debris disks over time. We used 100 Myr bins and fit a logarithmic profile to the decline of the number of debris disks as a function of age. We find that the number of debris disks declines as e$^{t0/t}$ where t$_{0}$$\sim$175 Myr.\label{fig4}}
\end{figure}

\clearpage
\begin{figure}
\center
\title{Dust Radii}
\includegraphics[width=150mm]{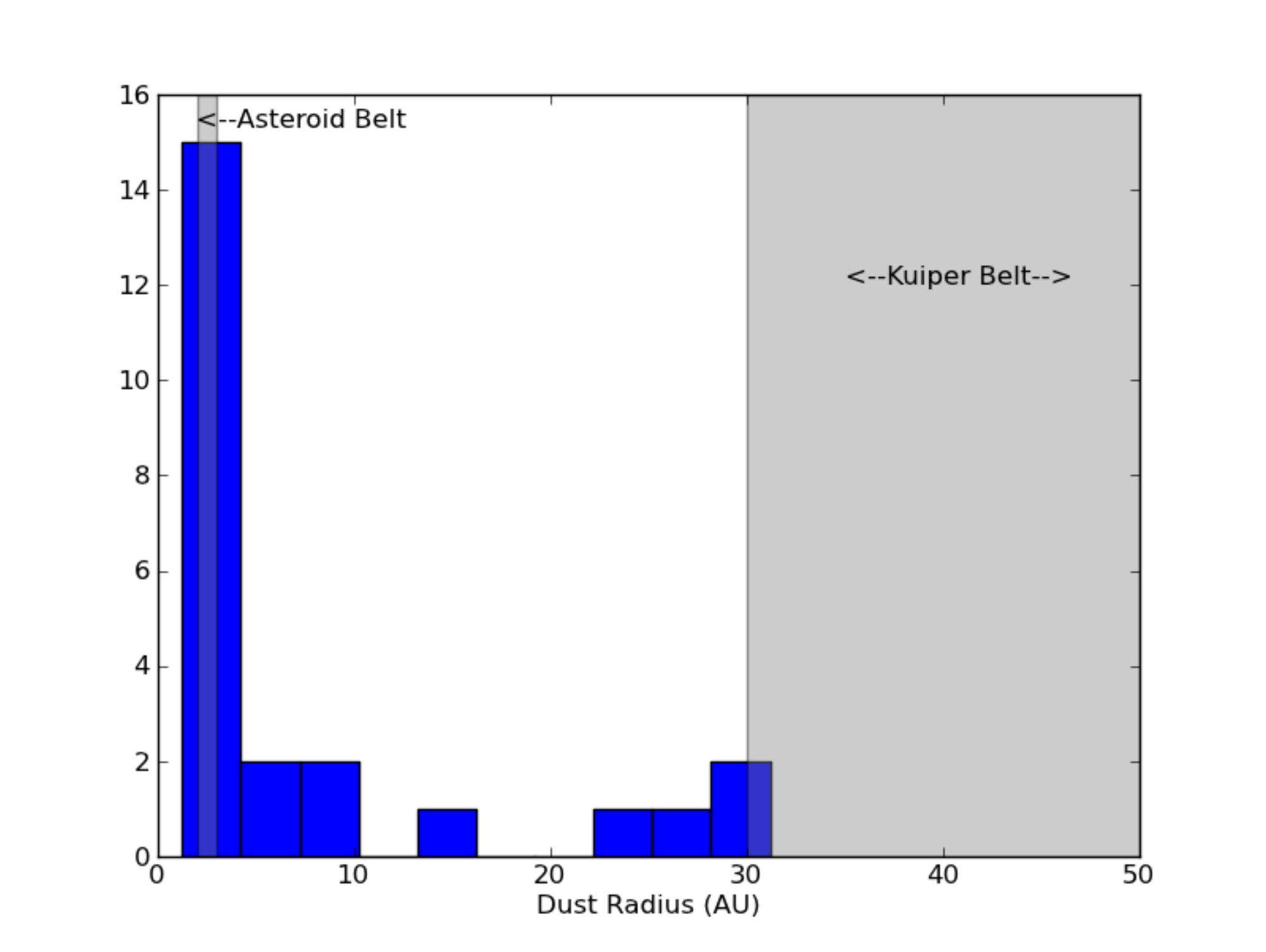}
\caption{This figure shows a histogram of dust radii for systems with dust blackbodies fit to more than one mid-IR data point. The locations of the Asteroid Belt and Kuiper Belt of the Solar System are shown for comparison, along with the orbits of the 4 giant planets.\label{fig5}}
\end{figure}
\clearpage

\begin{landscape}
\begin{figure}
\center
\title{Debris Disk Visualization}
\includegraphics[width=250mm]{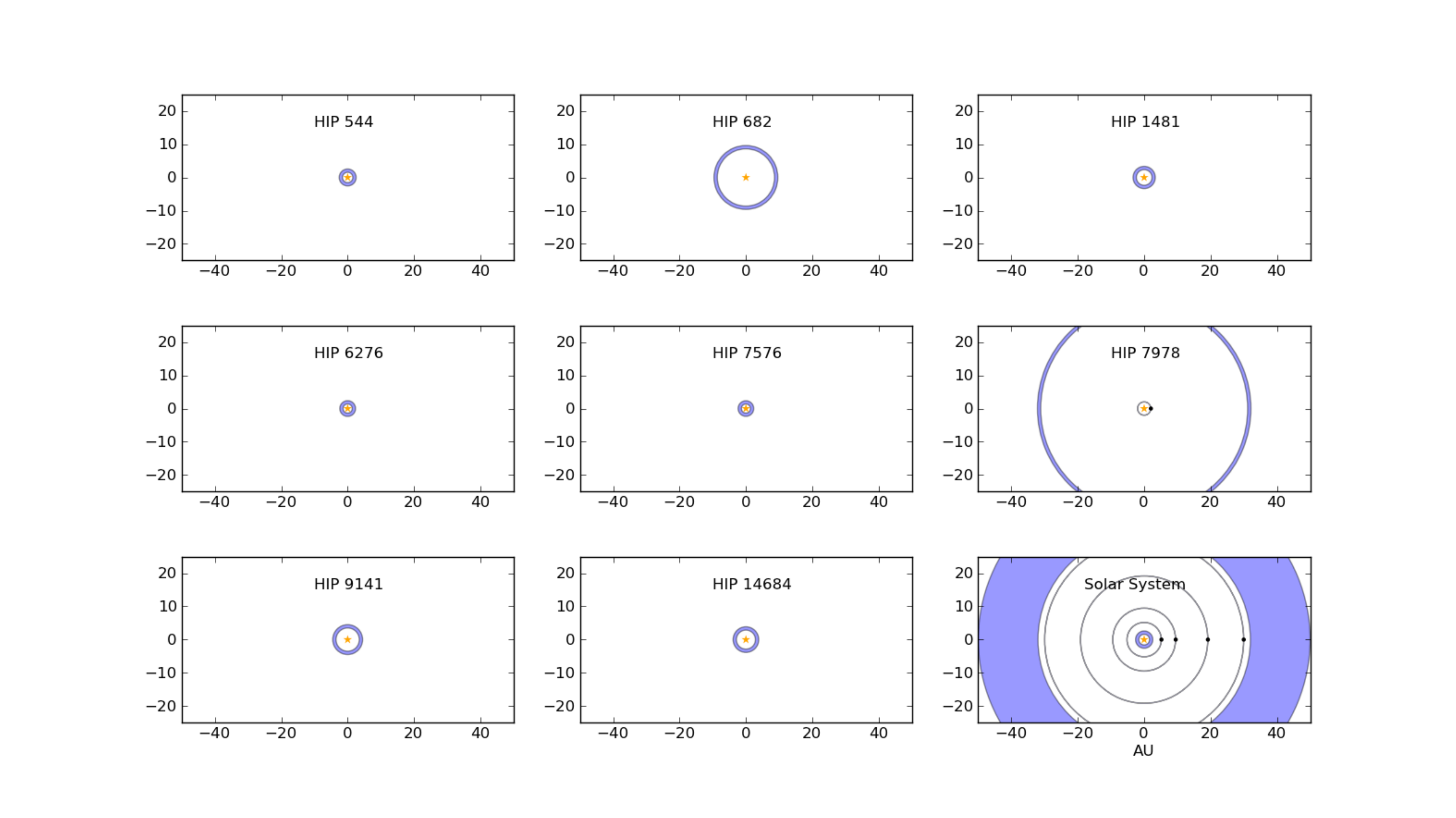}
\caption{This figure is a visualization of debris disks in systems with well determined dust radii. Note that the thickness of the dust rings is artificially added. Two stars are previously known planet hosts (HIP 7978 and HIP 118319).\label{fig6}}
\end{figure}
\end{landscape}
\clearpage

\begin{landscape}
\begin{figure}
\center
\title{Debris Disk Visualization (cont'd)}
\includegraphics[width=250mm]{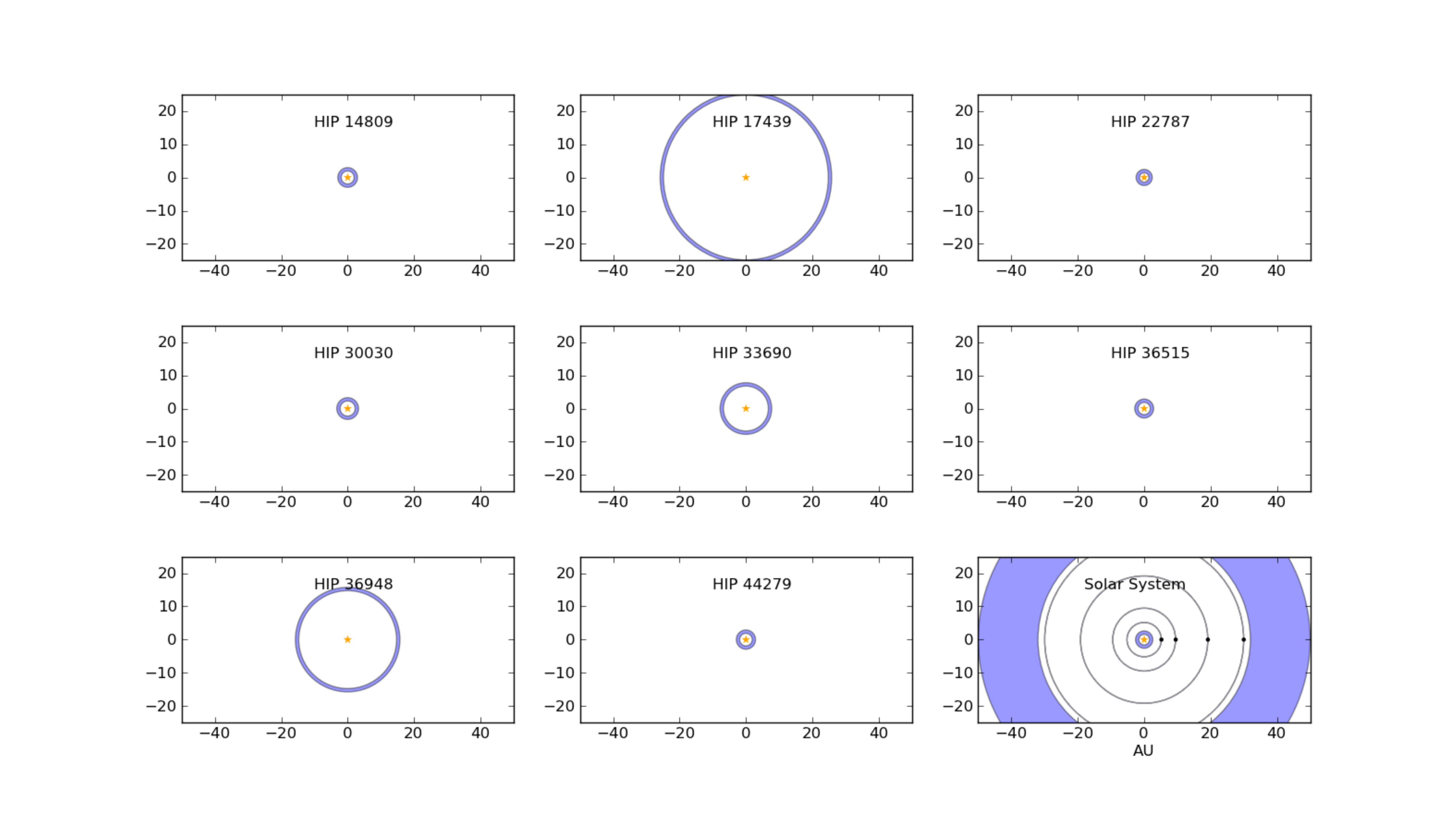}
\caption{Same as Figure 6.\label{fig7}}
\end{figure}
\end{landscape}
\clearpage

\begin{landscape}
\begin{figure}
\center
\title{Debris Disk Visualization (cont'd)}
\includegraphics[width=250mm]{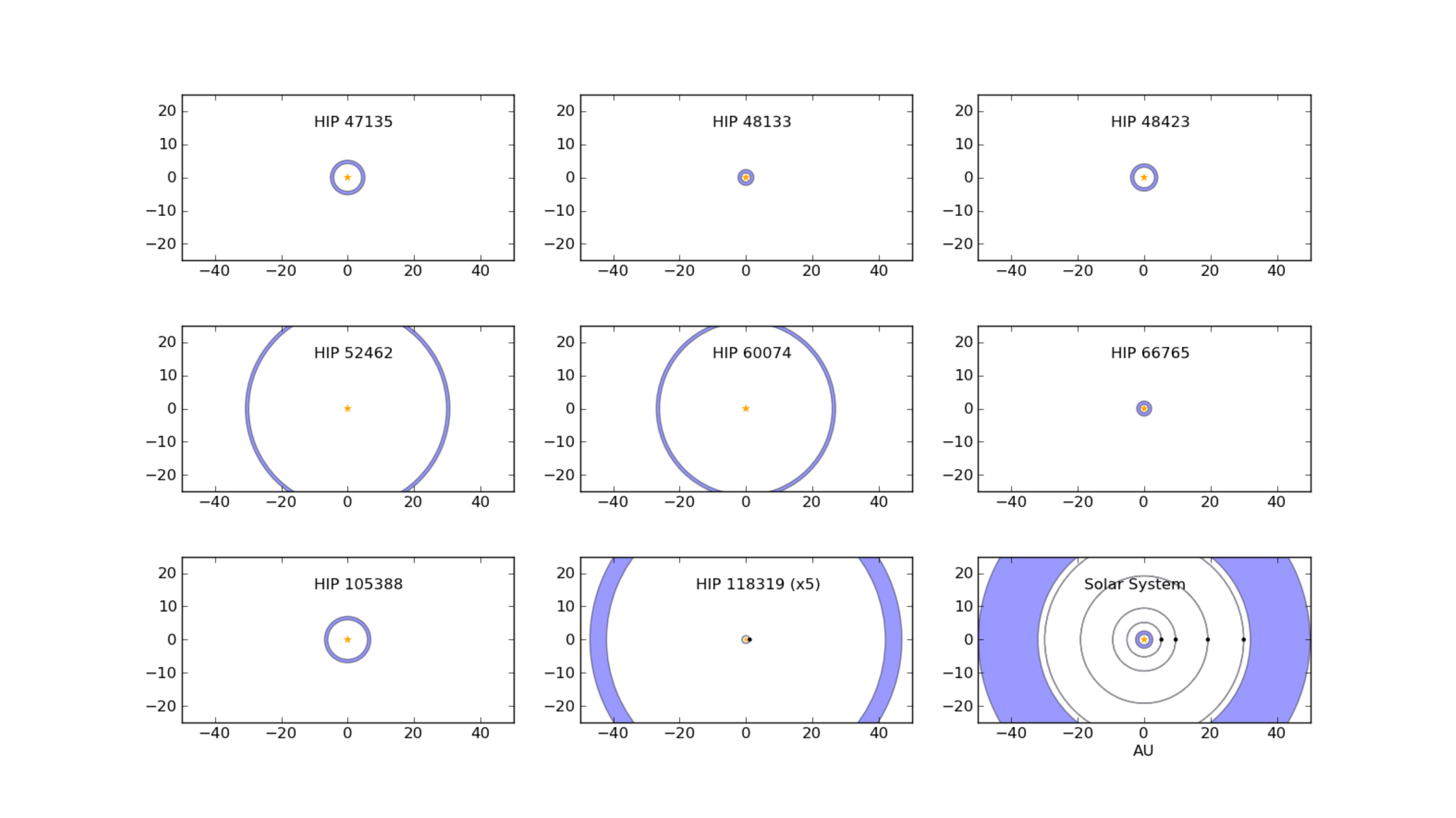}
\caption{Same as Figure 6. The scale of the debris disk and planet orbit for HIP 118319 is multiplied by 5 for easier visualization.\label{fig8}}
\end{figure}
\end{landscape}

%\clearpage
%\begin{figure}
%\center
%\title{Distribution of $\tau$}
%\includegraphics[width=150mm]{VicanFigure5.eps}
%\caption{The behavior of $\tau$, which traces the dust mass, over time. We reproduce a similar trend as has been observed in previous studies (e.g. Su et al. 2006, Bryden et al. 2006, Rhee et al. 2007). We present this figure with the caveat that most of our new debris disks only have one data point in excess - thus the blackbody fits represent the maximum temperature, and thus the minimum $\tau$ value. \label{fig5}}
%\end{figure}

\begin{figure}
\center
\title{Distribution of Distances}
\includegraphics[width=150mm]{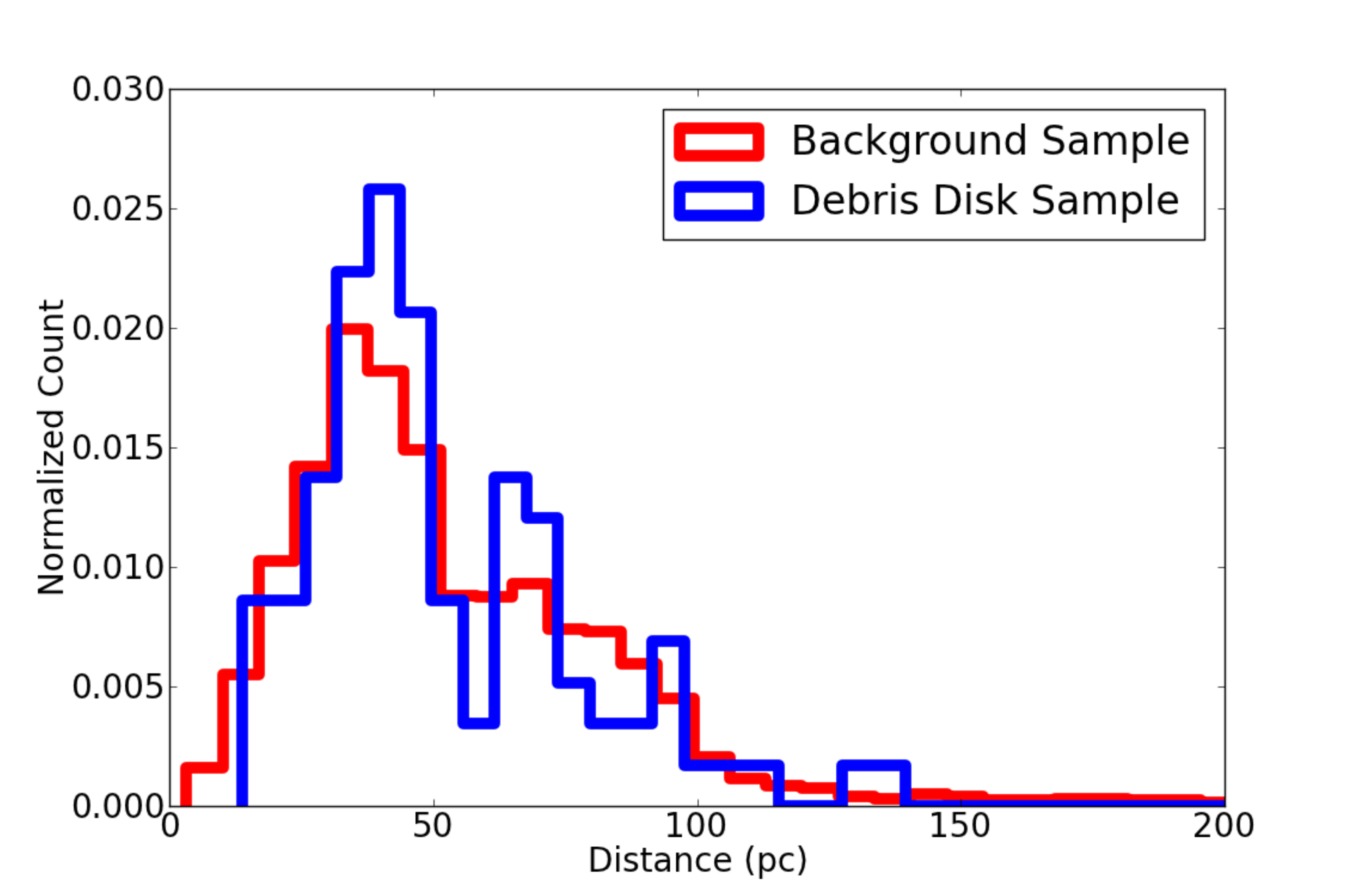}
\caption{The distribution of debris disks as a function of distance relative to the background sample (which includes the debris disk stars). It is clear that the frequency of debris disk detection has no dependence on distance (inside of 150 pc). Detected debris disks are too few to tell if there is any dependence on distance past 150 pc.\label{fig9}}
\end{figure}
\clearpage

\clearpage

\begin{figure}
\center
\title{Distribution of [Fe/H]}
\includegraphics[width=150mm]{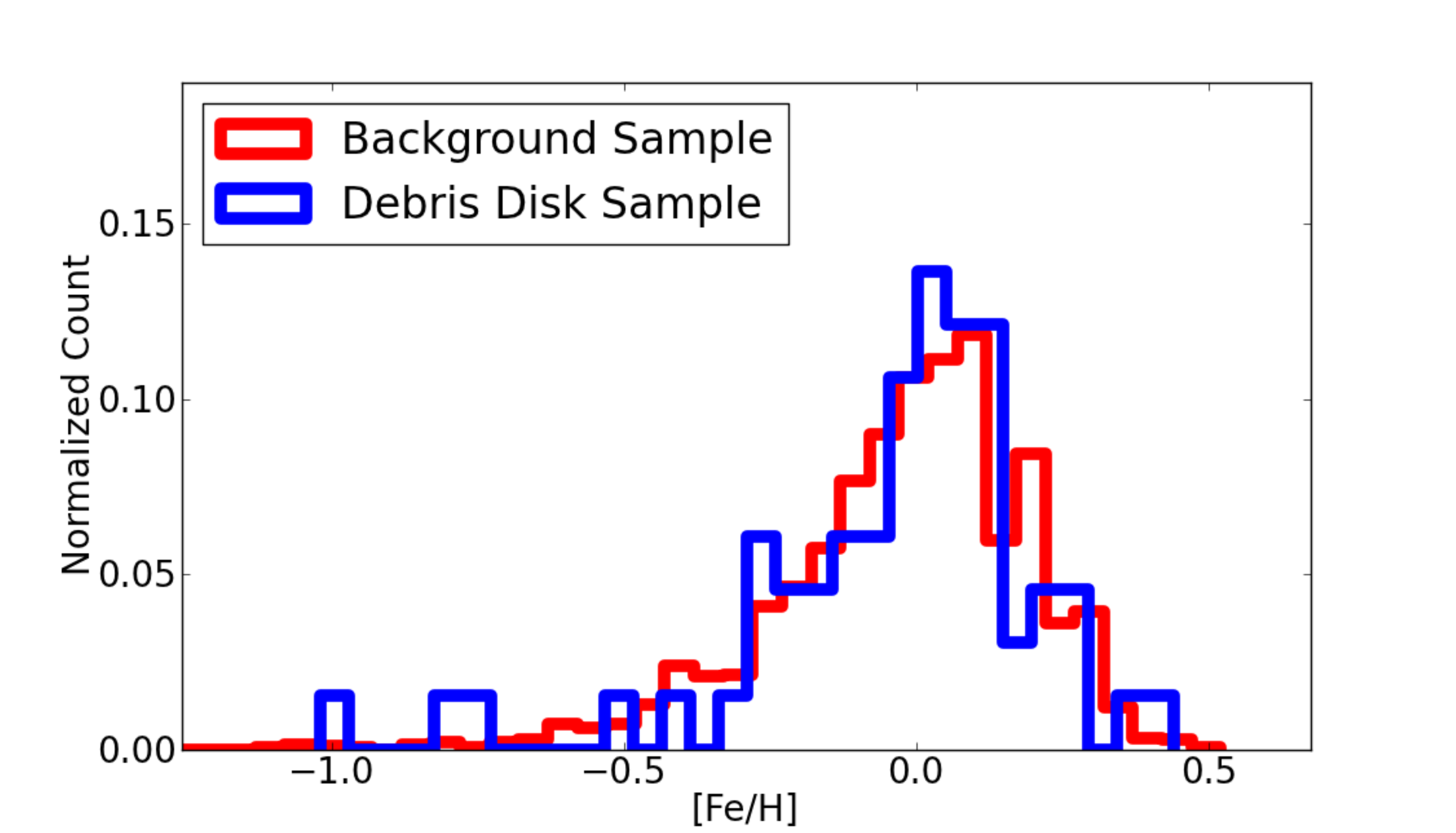}
\caption{This figure demonstrates the fact that debris disk incidence is not dependent on the metallicity of the host star. A similar trend was noticed by Greaves et al. (2005). This histogram includes 2,110 stars with measured metallicities from Anderson et al. (2012), 66 of which are debris disk hosts.\label{fig10}}
\end{figure}

\clearpage
%
%\begin{figure}
%\center
%\title{Age vs. Distance}
%\includegraphics[width=150mm]{VicanFigure8.eps}
%\caption{This figure demonstrates the fact that the ages of our sample do not depend on the distance from the Earth. The top panel represents stars from the entire background sample (including debris disks), while the bottom panel represents only stars with debris disks. It is also clear that this non-correlation is independent of spectral type.\label{fig8}}
%\end{figure}

\begin{figure}
\center
\title{Planet Hosts}
\includegraphics[width=150mm]{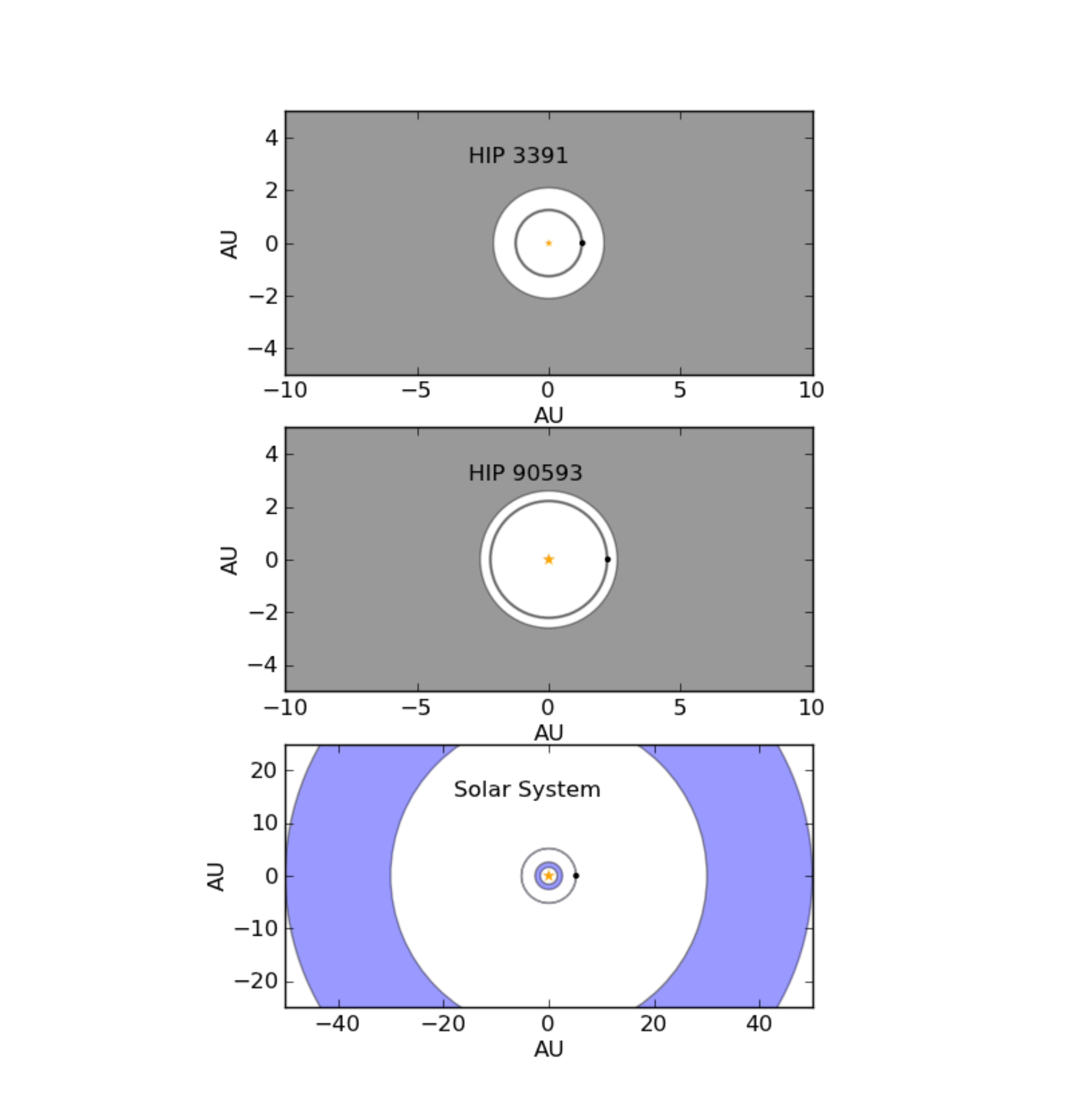}
\caption{This figure is a visualization of the debris disks around known planet hosts with poorly-determined dust radii. The disks around HIP 3391 and HIP 90593 are located at the minimum possible radius for these systems (corresponding to the maximum possible dust temperature). For comparison, we include the ``debris disks" of the solar system (the Asteroid and Kuiper Belts) with the orbit of Jupiter shown. \label{fig11}}
\end{figure}

\begin{figure}
\center
\title{Li Evolution}
\includegraphics[width=150mm]{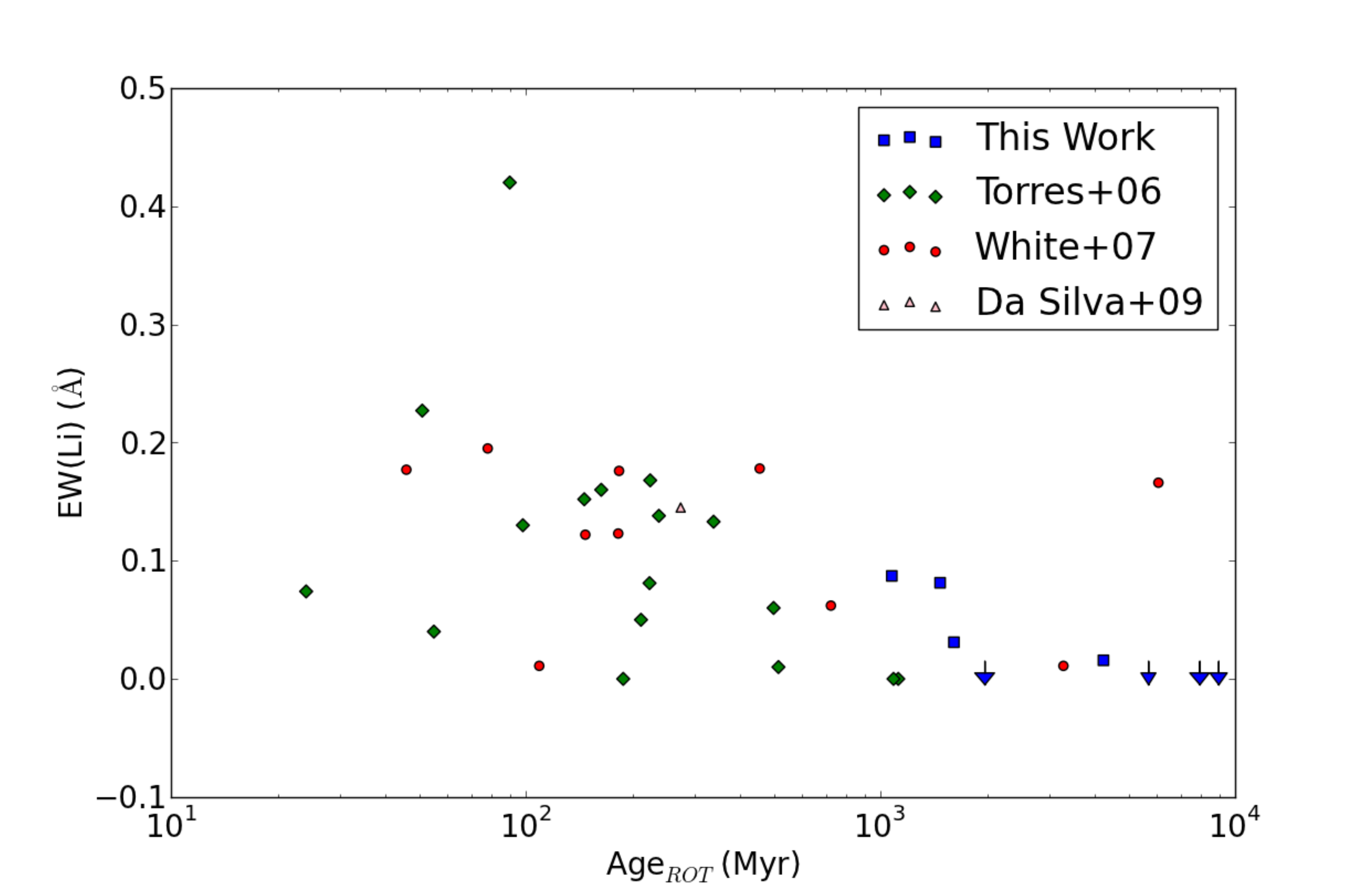}
\caption{Comparing lithium abundances (measured by the equivalent width (EW) of the Li absorption line at 6707.8 $\AA$) with the ages derived using the activity-rotation-age relation of MH08, we see that there is an apparent decrease in Li as a function of age. We expect the rate of this evolution to be mass-dependent (e.g. Zuckerman \& Song 2004); this figure conflates all F, G, and K-type stars. \label{fig12}}
\end{figure}

\clearpage

\begin{landscape}
\begin{figure}
\center
\title{Blackbody Degeneracy - HIP 90593}
\includegraphics[width=215mm]{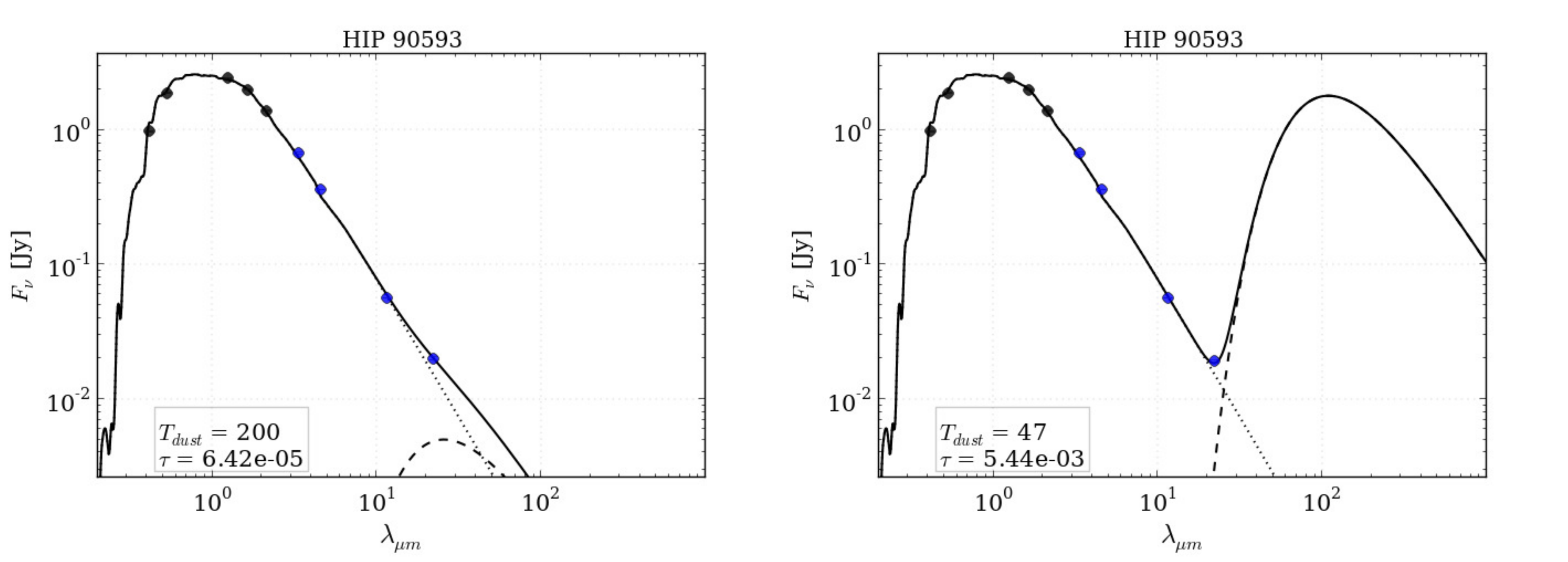}
\caption{Here we present two possible dust blackbodies which could be fit to the same excess point for HIP 90593 (one of our planet hosts). Clearly there is a large discrepancy in both the temperature and $\tau$ values derived from a single data point excess. Further observations are needed to constrain dust characteristics.\label{fig13}}
\end{figure}
\end{landscape}

%\begin{landscape}
%\begin{figure}
%\center
%\title{Comoving Pair HIP 9769 and HIP 9774}
%\includegraphics[width=215mm]{VicanFigure11.eps}
%\caption{HIP 9769 and HIP 9774 form a comoving pair (see Appendix A). Both stars have evidence of an excess at W4, and their old ages (t$\sim$8 Gyr) make them targets of particular interest.\label{fig11}}
%\end{figure}
%\end{landscape}

\end{document}